\documentstyle[12pt]{article}

\input epsf
\begin{document}\setlength{\unitlength}{1mm}

\def\question#1{{{\marginpar{\small \sc #1}}}}
\newcommand{\QCD}{{ \rm QCD}^{\prime}}
\newcommand{\MSSM}{{ \rm MSSM}^{\prime}}
\newcommand{\eq}{\begin{equation}}
\newcommand{\en}{\end{equation}}
\newcommand{\bino}{\tilde{b}}
\newcommand{\tsquark}{\tilde{t}}
\newcommand{\gluino}{\tilde{g}}
\newcommand{\photino}{\tilde{\gamma}}
\newcommand{\wino}{\tilde{w}}
\newcommand{\mtilde}{\tilde{m}}
\newcommand{\higgsino}{\tilde{h}}
\newcommand{\gsi}{\,\raisebox{-0.13cm}{$\stackrel{\textstyle>}
{\textstyle\sim}$}\,}
\newcommand{\lsi}{\,\raisebox{-0.13cm}{$\stackrel{\textstyle<}
{\textstyle\sim}$}\,}

\rightline{RU-96-25}
\rightline{hep-ph/9605266}
\rightline{May, 1996}
\baselineskip=18pt
\vskip 0.6in
\begin{center}
{ \LARGE High Temperature Dimensional Reduction of the MSSM and other Multi-Scalar Models}\\
\vspace*{0.6in}
{\large Marta Losada}\footnote{On leave of absence from Centro Internacional de
F\'{\i}sica and Universidad Antonio Nari\~{n}o, Sante Fe de Bogot\'a, COLOMBIA. }\\
\vspace{.1in}
{\it Department of Physics and Astronomy \\ Rutgers University,
Piscataway, NJ 08855, USA}\\
\end{center}
\vspace*{1.0in}
\vskip  0.4in  

Abstract: Using dimensional reduction we construct an effective 3D theory of the
 Minimal Supersymmetric
Standard Model at finite temperature. The final effective theory is obtained
after three successive stages of integration out of massive particles. We obtain the
full 1-loop relation between the couplings of the reduced theory and the underlying
4D couplings and masses. The
procedure is also applied to a general 
two Higgs doublet model and the Next to Minimal Supersymmetric Standard Model.

\thispagestyle{empty}
\newpage
\addtocounter{page}{-1}
\newpage

\section{Introduction}

\hspace*{2em} A crucial problem in particle physics is the understanding of the observed
baryon asymmetry of the universe. A possible scenario which has been
explored involves physics at the electroweak scale. For electroweak
baryogenesis to occur via a sufficiently strong first order phase transition,
such that  baryon number violation is suppressed after the phase transition,
 the sphaleron transition rate in the
low temperature phase must be less than the universe expansion rate \cite{sakharov}.
This requires that the ratio of the vacuum expectation value in the broken phase
to the phase transition temperature $(T_{c})$ must be \cite{shaposhnikov}

\begin{equation}
{v(T_{c})\over T_{c}} \gsi 1,
\label{vevTc}
\end{equation}
which in turn
places constraints on the Higgs structure of the low temperature theory. In fact,
 these requirements may not be fulfilled  in the 
Standard Model 
at least for experimentally allowed values of the Higgs mass \cite{shaposhnikov, dine}.

 In order to study the
properties of the phase transition, standard perturbation techniques (with resummation)
were initially  utilized. However, there are two intrinsic problems with perturbation
theory. First of all, the loop expansion parameter ($ {g^{2} T\over M_{W}} \sim {\lambda\over g^{2}}$) is proportional to the
 zero temperature
ratio of Higgs mass to the W boson mass.  This is implies that unless
the Higgs is sufficiently light the perturbative approximations will break down
when higher order corrections are included \cite{espinosa}. 
Secondly, high temperature gauge theories
 are subject to infrared divergences arising from massless gauge boson modes in the theory \cite{gross}. 
In the broken phase these divergences are absent because the gauge bosons of the broken 
symmetry have non-zero $\phi$-dependent masses.
Consequently, even when the aforementioned
 perturbative effects are under control, non-perturbative physics in the unbroken phase, for values
of $\phi \sim 0$, can limit the reliability of the
perturbative calculations  of the phase transition. In other words,  
knowledge of the effective potential for small values of the
scalar field  is required to study the properties of the phase transition, and
the assumptions underlying perturbative calculations must be verified by
calculations which include non-perturbative effects.

Using the fact that in the symmetric phase at finite temperature the long distance
theory is described by a three dimensional bosonic
gauge theory, Kajantie, Laine, Rummukainen, Shaposhnikov and Farakos \cite{farakos, KLRS, kajantie}
devised a procedure which separates the perturbative and non perturbative aspects of the
study of the electroweak phase transition. As explained in section 2.1,  due to antiperiodicity
of the boundary condition
at finite temperature,  fermions in the theory acquire thermal masses proportional
to the temperature $\sim \pi T$. The bosonic field decomposition into thermal
modes contains a static mode which does not have a thermal mass, as well as modes with masses  on the
order of $\sim \pi T$. 
 The construction of the effective theory for the static modes amounts to
perturbatively integrating out the effects of all the massive modes. As a consequence
of the interaction with the heavy modes,
the masses of some of  the static scalar fields are modified enough that they 
become heavy ($\sim gT$) and one can further construct a second effective 3D theory describing
only the light fields. This reduced three
dimensional gauge theory for the light scalars can then be analyzed by numerical  (lattice) calculations so that
all non perturbative effects are handled correctly.

 Kajantie et al.  reduced the Standard Model to a 3D gauge theory with a single
light scalar, which they studied on the lattice. They concluded that many, but not
all,  features of the phase transition were similar to results of perturbation theory. The results
showed
that there is no value of the Higgs mass in the Minimal Standard Model which is compatible 
with correct electroweak symmetry breaking in the low temperature theory, given the measured
large top mass \cite{KLRS}.

Consequently, it is of interest to analyze extensions of the Standard Model to see whether
there are any cases in which there is a sufficiently strong first order phase transition.
Using the numerical results of Kajantie et al. we can include non perturbative effects
for any 4D theory which reduces to an effective three dimensional theory containing a single light scalar coupled
to $SU(2)$ gauge fields. This is the generic situation at the
phase transition, except when the theory is fine-tuned, even when the theory
has multiple scalar fields as in supersymmetry, general two Higgs doublet models, etc.

A feature of multi-scalar models is the inclusion of tree level masses
for all scalars. Additionally, some of the scalar fields
may carry colour charge. This implies that in the unbroken phase the static modes of the scalars
can acquire masses between $\pi T$ and $gT$. In particular, the masses of the squarks static modes
receive corrections proportional to $\sim g_{s} T$. Consequently, unlike the Standard Model case,
 we encounter an additional intermediate effective theory in which the static modes 
with masses in this range
are integrated out before the final reduction procedure of modes with masses $\sim gT$ is performed.
For supersymmetric theories the static
squark (slepton) modes are integrated out at this second stage. 

For the generic case, the reduction procedure leaves us with the same effective
theory as in \cite{KLRS}, defined by the effective
couplings $\bar{\lambda}_3$ and $g_3^2$ with a
different relationship which define these quantities in terms of the masses and
couplings of the 4D theory.  As explained by Kajantie et al \cite{kajantie},
the dynamics of the electroweak phase transition
is governed by the quantity, 

\begin{equation}
x_c={\frac{\bar{\lambda}_3}{g_3^2}}
\label{xcrit}
\end{equation}
 at the critical temperature, and the constraint given by equation (\ref{vevTc}) translates into \cite{KLRS}

\begin{equation}
x_{c} < 0.04.
\label{xmax}
\end{equation}

In this paper we present the  relations between the
masses and couplings of the reduced theory and the  parameters  of the 
underlying four dimensional theory for three
 models: the Minimal Supersymmetric Standard Model (MSSM), a general two Higgs doublet
model (2HDM) and the Next to Minimal Supersymmetric Standard  Model (NMSSM). The analysis
of the electroweak phase transition for the MSSM will be the subject of a second paper \cite{ewptmssm}.
In section 2.1 we review the finite temperature formalism which is the
 basis for the construction
of the three dimensional  theory. Sections 2.2, 2.3 and 2.4 discuss the
three effective theories for the MSSM as we
 integrate out in
successive stages, (1) the non-static modes, (2) the heavy squarks and sleptons and (3) the
heavy Higgs and the temporal component of the $SU(2)$ gauge field $A_o$. Section 2.5 discusses
the non-generic case in which there are two light scalar fields in the
final effective theory. We present
a short discussion of other multi-scalar models in Section 3.
 Our conclusions  are given in Section 4. Appendix A introduces the 
MSSM 4D Lagrangian, 
which is the model discussed in the text. Appendices B.1 and B.2 give the 
explicit expressions for our results of the first and second stage. 
In appendix B.3 we give the results for the  case in which there are two light scalar 
Higgs fields in the final effective theory. The results of the application
of the procedure to the two Higgs doublet
model and the NMSSM is given in appendices C.1 and C.2. Finally, we present
some useful high temperature formulae in appendix D.

\section{Three Dimensional Theory}

\subsection{Effective Theory}

\hspace*{2em} A  given theory in four dimensions at finite temperature reduces,
in the high temperature limit, to an
effective 3D theory describing the static degrees of
freedom. This is the statement of dimensional reduction \cite{applequist,
nadkarni, kajantie, landsmann}.

The action at finite temperature is given by

\begin{equation}
S = \int_o^{\beta} d\tau \int d^{3}x  \cal{L},
\label{action}
\end{equation}
where $\beta = {1\over T}$, such that the theory is characterized by bosonic (fermionic)
fields obeying periodic (anti-periodic) boundary conditions in Euclidean
time \cite{kapusta}. The finite temperature expansion of the fields in our
theory is given by

\begin{equation}
{\cal S} (x, \tau) = T^{1/2} [ {\cal S}_{0}(x) + \sum_{n\neq 0} {\cal S}_{n} (x) e^{i 
\omega_{B} \tau}]
\label{phi13d}
\end{equation}

\begin{equation}
\Psi (x, \tau) = T^{1\over 2} \sum_{n} \psi_{n} e^{i\omega_{F}\tau} ,
\label{fermions}
\end{equation}
where ${\cal S}$ and $\Psi$ represent bosonic and fermionic fields, respectively.
$ {\cal S}_{0}$ is the static component of the bosonic field, for which $\omega_{B}=0$.
The resulting propagators are of the form $(k^2+\omega
_{B(F)}^2+m^2)^{-1}$, where $\omega_B=2\pi nT,\omega_F=\pi (2n+1)T$ for
bosons and fermions, respectively. That is, non-static modes have masses
$\sim \pi T$ and thus all of the fermionic
modes may be integrated out at large $T$.
Our calculations will be carried out in Landau gauge.  By integrating over $\tau $
and using the orthonormality of the modes we can obtain the terms in the
Lagrangian describing the static modes, the non-static modes and finally the interaction
terms between the heavy and light modes. 

\subsection{First Stage}

\hspace*{2em} We describe in detail the dimensional reduction of the MSSM. In the appendices we treat a general 2HDM
model and supersymmetric models with an additional gauge singlet superfield. We work in the
$\sin^{2} \theta_{W} =0$ limit and ignore the hypercharge $U(1)$ gauge boson 
and gaugino. The Lagrangian for the MSSM is given in Appendix
A, where we introduce our notation. Most of the discussions in the paper will be general enough to include all
three models. We leave the notation and particular results  corresponding to the 2HDM
and NMSSM to Appendix C. 
\vspace{.2in}

Starting with the 4D theory we generalize the procedure delineated by
Kajantie et al. to obtain a 3D effective theory.
The first stage corresponds to the integration of the massive (non-static)
modes. As a result, all fermions are integrated out and we are left with a
three dimensional bosonic theory. We can systematically construct the dimensionally
reduced theory and relate its parameters to those of the more fundamental four dimensional
theory by computing the effective interactions among the static modes generated by integrating
out the $n\neq 0$ modes. This procedure is perturbative in the coupling constants of the theory and we
will calculate consistently throughout to order $g^{4}$. As explained in \cite{ KLRS, kajantie}, for our purposes it it 
sufficient to perform a one loop calculation as it provides the specified accuracy and  the determination
of the critical
temperature is precise enough as 
 the temperature dependence in the ratio in equation (\ref{xcrit}) enters only  through
logarithmic terms.

As a result of the first stage of integration the $n=0$ bosonic modes in the theory acquire a thermal mass
to leading order (excluding logarithmic corrections) of the generic form:

\begin{equation}
m^2(T) = m^2 + \nu T^{2} ,
 \label{Tmass}
\end{equation}
 here $m^{2}$ is the tree level mass squared, which for gauge bosons is identically
zero. As explained above the quantity $\nu$ is determined from the one loop integration and
in particular it
remains zero for the spatial components of the vector field  $A_{i}$. This reflects the fact that the $A_{i}$ 
fields are precisely the gauge fields of the 3D theory. For the
temporal component of the $SU(2)$ gauge field $A_{o}$, which is a gauge-triplet of scalars in the
effective theory, $\nu \neq 0$. This implies that the temporal mode acquires a mass, the so-called Debye mass,
which will be on the order of $\sim g T$.\footnote{Similarly the longitudinal $SU(3)$ gauge
field acquires a mass  $\sim g_{s} T$.}
Scalar particles, on the other hand,  have a  tree level mass as well as a non-zero value of 
$\nu$. In the high temperature approximation, their masses are of order
 $(m^2 + g_{s}^{2} T^{2})^{1\over 2}$ for squarks. For stops the contribution 
proportional to the top Yukawa coupling squared can also be significant.
 The rest of the scalars in the theory will have masses on the order of $(m^2 + g^{2} T^{2})^{1\over 2}$.
Higher order effects are suppressed by powers of $m^{2}/T^{2}$. 

 The exact value of the
tree level mass is important as it may lead to the existence of a very light scalar particle.
 This may have several 
different consequences but specifically, if the tree level mass nearly cancels the 
term proportional to $T^{2}$ one could be close to a phase transition. We will 
discuss the non-generic case in which there are two light scalar
particles in the final effective theory in section 3.

In order to establish our notation we write the 3D scalar potential after the
first stage,

\begin{eqnarray}
\lefteqn{V(A_{o},\phi_{1}, \phi_{2}, Q_{i}, U_{i}, D_{i}) = {1\over 2} M_{D}^{2} A_{o}A_{o} + H(A_{o}A_{o})
 (\phi_{1}^{\dagger}\phi_{1} + \phi_{2}^{\dagger}\phi_{2}
 + \sum_{i} Q_{i}^{\dagger}Q_{i})} \nonumber \\
&+& M_{1}^{2} \phi_{1}^{\dagger}\phi_{1} + M_{2}^{2} \phi_{2}^{\dagger}\phi_{2} 
+ M_{3}^{2} (\phi_{1}^{\dagger}\phi_{2} + \phi_{2}^{\dagger}\phi_{1})
+ \Lambda_{1} (\phi_{1}^{\dagger}\phi_{1})^{2} +
\Lambda_{2} (\phi_{2}^{\dagger}\phi_{2})^{2}  \nonumber \\
 &+& \Lambda_{3}
 (\phi_{1}^{\dagger}\phi_{1})(\phi_{2}^{\dagger}\phi_{2}) +
\Lambda_{4} (\phi_{1}^{\dagger}\phi_{2})(\phi_{2}^{\dagger}\phi_{1}) + \sum_{i}\Lambda_{3}^{Q_{i}1}
 (\phi_{1}^{\dagger}\phi_{1})( Q_{i}^{\dagger}Q_{i}) \nonumber \\
&+& \Lambda_{3}^{Q_{i}2}
 (\phi_{2}^{\dagger}\phi_{2})( Q_{i}^{\dagger}Q_{i})
 + (\Lambda_{4}^{Q_{i}1} + \bar{f_{d_{i}}^{L}}^{2})
 (\phi_{1}^{\dagger}Q_{i})( Q_{i}^{\dagger}\phi_{1})
 + (\Lambda_{4}^{Q_{i}2}  \nonumber \\
&+& \bar{f_{u_{i}}^{L}}^{2})|\epsilon_{\alpha\beta}
\phi_{2}^{\alpha} Q_{i}^{\beta}|^{2}
+ M_{Q}^{i}  Q_{i}^{\dagger}Q_{i}+ M_{u}^{i} U_{i}^{\dagger}U_{i}
 + M_{d}^{i} D_{i}^{\dagger}D_{i} + \bar{f_{d_{i}}^{R}}^{2} 
(\phi_{1}^{\dagger}\phi_{1})( D_{i}^{\dagger}D_{i}) \nonumber \\
&+&  \bar{f_{u_{i}}^{R}}^{2} 
 (\phi_{2}^{\dagger}\phi_{2})( U_{i}^{\dagger}U_{i}) 
+ \bar{A} f_{d_{i}}\phi_{1}^{\dagger} Q_{i} D_{i}
- \epsilon_{\alpha\beta} \bar{A} f_{u_{i}}\phi_{2}^{\alpha} Q_{i}^{\beta} U_{i}
+ \bar{\mu} f_{d_{i}}\phi_{2}^{\dagger} Q_{i} D_{i}^{\ast}\nonumber \\
&-&\epsilon_{\alpha\beta} \bar{\mu} f_{u_{i}}\phi_{1}^{\alpha} Q_{i}^{\beta^{\ast}} U_{i}^{\ast} + h.c.
\label{3dscalarpot}
\end{eqnarray}
where $\alpha$, $\beta$ are $SU(2)$ indices. Note that the fields
in equation (\ref{3dscalarpot}) are the static components of the scalar fields, properly
renormalized, and  having dimension [GeV]$^{1\over 2}$. Quartic couplings have dimensions
of [GeV] and trilinear couplings  involving $\bar{A}$ or $\bar{\mu}$ have dimension of [GeV]$^{3\over 2}$. We have omitted the terms corresponding 
to the scalar leptons and the quartic $A_{o}$ term.

The full expressions for the effective masses and couplings are given in Appendix B.1 where
we also show the diagrams contributing to the calculations for each one of the parameters.
We would like to point out some of the features of the calculations, though most technical
details are discussed in Appendix B.

The expressions for the 3D parameters  contain temperature dependent logarithmic
corrections denoted 
$L_{b}$ and $L_{f}$.  In general the coefficients of these quantities are the
 corresponding bosonic and fermionic contributions to the beta functions
for the masses and couplings in the underlying theory \cite{farakos}. 
In the 4D theory the
 quartic scalar couplings (Higgs self couplings 
and part of the interactions between Higgses and squarks/sleptons), are fixed
above the SUSY breaking scale in terms of the gauge couplings. In the zero temperature theory,
at  energy scales above the
SUSY breaking scale, or if there is no
particle decoupling, the beta function coefficients
are related by the same algebraic relation as the couplings themselves \cite{hempfling}.
 This is no longer true if one considers particles 
decoupling in the 4D theory below the SUSY scale.
Similarly, in the  finite
temperature theory  the relation between the (scale dependent part of the)
3D expressions of the couplings differ, upon
1-loop integration of heavy modes, from the expression relating the
beta functions  of the zero temperature theory.  As a by-product our 3D expressions for the
couplings and masses, yield the full 1-loop beta function coefficients
in the zero  temperature theory, including particle decoupling for the three models we have considered.

The static modes squark (slepton) masses are non-degenerate as a result of the
integration procedure even if their masses are taken to be
degenerate in the 4D theory. Not only do right and left handed type squarks acquire
different values for their masses, so do up and down type right handed
squarks. This occurs because right handed squarks do not couple 
to $SU(2)$ gauge fields, their quartic
coupling to Higgses is proportional to their corresponding Yukawa
coupling, and the trilinear coupling to Higgses differs for up and
down type right handed squarks.

As in the case of the Standard Model, the gauge coupling between the spatial magnetic field
and scalars in the three dimensional theory is not equal to the quartic coupling between the
$A_{o}$ field and scalars. In addition, at the next stage of 
integration-out this latter quartic scalar coupling will  be different
for each type of scalar field to which the $A_{o}$ couples to as a consequence of the
soft SUSY breaking trilinear couplings.

\vspace{.3in}
\subsection{Second Stage}

\hspace*{2em} The aim of the second and third stage is twofold. First, as explained 
qualitatively above, after the first stage of integration we are still left with
several different mass scales, while the purpose of an effective theory is
to have only one characteristic mass scale. Secondly, if we can construct an  effective
theory in which we are left with only one light scalar particle then we have arrived at 
exactly the same theory which has already been analyzed on the lattice by Kajantie et al \cite{KLRS}.

What we define as  second stage is necessary only when the  mass of the squarks
and sleptons is such that the high temperature expansion is valid. If the squarks and
sleptons were extremely massive they would have decoupled in the four dimensional
theory, or alternatively a low temperature expansion might be applicable \cite{hall}.
After obtaining our reduced theory  we must verify that the non-renormalizable terms of the
effective theory are indeed suppressed. That is, we must check
 that higher order corrections to the scalar potential at the critical temperature
do not change qualitatively our results.

For the MSSM and NMSSM the second stage corresponds to the integration of heavy squarks and
sleptons. We include the sleptons even though their masses do not have contributions $\sim g_{s} T$, since
their tree level mass at some high scale is presumably $\gsi gT$. The results are also
applicable to  the case in which
we study a purely $SU(2)$ gauge theory with multiple scalars, if we ignore all of the contributions
which include $g_{s}$.
 For the MSSM the resultant theory  after the second stage is described by a 2HDM 
 with complicated expressions, in terms of
the 4D couplings and masses,  for the masses and strengths of
interactions. The 3D potential for the scalar fields $A_{o}$, $\phi_{1}$
and $\phi_{2}$ is

\begin{eqnarray}
V(A_{o},\phi_{1}, \phi_{2})& = & {1\over 2} \bar{M_{D}^{2}} A_{o}A_{o} + \bar{H_{1}}(A_{o}A_{o})
 \phi_{1}^{\dagger}\phi_{1} + \bar{H_{2}}(A_{o}A_{o})
 \phi_{2}^{\dagger}\phi_{2} \nonumber \\
&+& \bar{M_{1}^{2}} \phi_{1}^{\dagger}\phi_{1} + \bar{M_{2}^{2}}
 \phi_{2}^{\dagger}\phi_{2} 
+ \bar{M_{3}^{2}} (\phi_{1}^{\dagger}\phi_{2} + \phi_{2}^{\dagger}\phi_{1})
+ \bar{\Lambda}_{1} (\phi_{1}^{\dagger}\phi_{1})^{2} \nonumber \\
& + &
\bar{\Lambda}_{2} (\phi_{2}^{\dagger}\phi_{2})^{2} + \bar{\Lambda}_{3}
 (\phi_{1}^{\dagger}\phi_{1})(\phi_{2}^{\dagger}\phi_{2}) +
\bar{\Lambda}_{4} (\phi_{1}^{\dagger}\phi_{2})(\phi_{2}^{\dagger}\phi_{1}) ,
\label{second3dpot}
\end{eqnarray}
where we have labeled the couplings and masses after the second stage with bars 
and for simplicity we maintain the same notation for the fields. The explicit
expressions for the parameters are given in Appendix B.2.

\subsection{Third Stage}

\hspace*{2em} After the second stage the scalar fields we are left with are  the two Higgs doublets and the
$A_{o}$ triplet. For a phase transition
to occur at least one of the thermal masses of the Higgses must become zero and
then negative as the temperature decreases. 
At this temperature we can determine the mass of the other Higgs doublet;
if it is heavy $\sim g T$,
it can be integrated out in a third stage, together with the $A_o$ field.

To determine which is the correct scalar Higgs field which is heavy at the
critical temperature one can analyze the eigenvalues of the mass matrix as a
function of the temperature. We find that only with fine-tuning  can one have two light
Higgs fields  in addition to the spatial gauge fields.

The critical temperature $T_{c}$, for which only one of the eigenvalues of the
mass matrix is zero is determined from
the following equation,

\begin{equation}
\bar{M_{1}}^{2}(T_{c})\bar{M_{2}}^{2}(T_{c}) = (\bar{M_{3}}^{2})^{2}(T_{c}).
\label{Tc}
\end{equation}
 Upon diagonalization of the scalar mass matrix the expression for the
mass of the
heavy Higgs field at the critical temperature is,

\begin{equation}
\nu^{2}(T_{c}) = \bar{M_{1}}^{2}(T_{c}) + \bar{M_{2}}^{2}(T_{c}).
\label{nuTC}
\end{equation}

  We denote by $\alpha_{i}$, the quartic Higgs coupling interactions in the second stage after 
diagonalization to the mass eigenstate basis requiring
one light Higgs. Where $\alpha_{1}$ is the quartic self-coupling of
 the massless Higgs field and
$\alpha_{3}$, $\alpha_{4}$ are quartic couplings between the light and 
heavy  Higgs scalar fields,

\begin{equation}
\alpha_{1} = \bar{\Lambda}_{1}\cos^{4}\theta + \bar{\Lambda}_{2}\sin^{4}\theta 
+ (\bar{\Lambda}_{3} + \bar{\Lambda}_{4})\cos^{2}\theta\sin^{2}\theta
\label{alpha1}
\end{equation}

\begin{equation}
\alpha_{3} = (2\bar{\Lambda}_{1} + 2\bar{\Lambda}_{2})\cos^{2}\theta\sin^{2}\theta + 
 \bar{\Lambda}_{3}(\cos^{4}\theta +
\sin^{4}\theta ) - 2\bar{\Lambda}_{4}\cos^{2}\theta\sin^{2}\theta 
\label{alpha3}
\end{equation}

\begin{equation}
\alpha_{4} = (2\bar{\Lambda}_{1} + 2\bar{\Lambda}_{2})\cos^{2}\theta\sin^{2}\theta 
 - 2\bar{\Lambda}_{3}
\cos^{2}\theta\sin^{2}\theta + \bar{\Lambda}_{4}(\cos^{4}\theta -
\sin^{4}\theta )
\label{alpha4}
\end{equation}
and

\begin{equation}
\tan 2\theta = {2\bar{M}_{3}^{2}\over (\bar{M}_{1}^{2} - \bar{M}_{2}^{2})}.
\label{theta}
\end{equation}

We now proceed to the third stage in which we integrate out the massive scalars 
which are left in the theory.\footnote{We have explicitly checked that
the precise order of integration out of the $A_{o}$ field, before or after diagonalization, 
is not relevant up to terms $\sim g^{6}$.}
 We are then left with an expression for the strengths of the interactions
of the static magnetic fields and the light Higgs field, in terms of
the quantities of the previous stage. In particular, we obtain the expression
 for the effective 3D
gauge coupling

\begin{equation}
g_{3}^{2} = \bar{G}^{2} \left(1 - {\bar{G}^{2}\over 24\pi}({1\over\nu(T_{c})} +
{1\over \bar{M_{D}}})\right),
\label{g32}
\end{equation}
and the effective Higgs self-coupling

\begin{equation}
\bar{\lambda}_{3} = \alpha_{1} - (\alpha_{3}^{2} + {\alpha_{4}^{2}\over 2} +
\alpha_{3} \alpha_{4}){1\over 8\pi\nu(T_{c})} - {3 (\bar{H_{1}}\cos^{2}\theta +
\bar{H_{2}}\sin^{2}\theta)^{2}\over 8\pi\bar{M}_{D}},
\label{lambda3bar}
\end{equation}
 in terms of
the original 4D parameters and the temperature. We point out that there is no
wavefunction renormalization at the third stage as there is no trilinear coupling
between the light Higgs and the heavy scalars.

If any of the squark (slepton) thermal masses were of the order of $\overline{M}_{D}$
rather than $>> \overline{M}_{D}$ they
would be integrated out at this point instead of previously. 
If, for example, we suppose the
sleptons to be light the modifications 
to the second stage equations would be such that  the sums in the equations
of Appendix B.2 would not run over the sleptons. Furthermore, at the third stage
after the rotation to the mass eigenstate basis the sleptons could
be integrated out together with the heavy Higgs and the $A_{o}$. The results of
the third stage, equations (\ref{g32}), (\ref{lambda3bar})
would contain additional terms of the same form as the contributions in the
second stage. However, it is clear that the expressions for the $\alpha_{i}$,
$\bar{G}^{2}$, $\bar{M}_{D}$, $\bar{H_{1}}^{2}$,  $\bar{H_{2}}^{2}$, $\nu(T_{c})$ would in
general be different.
We point out that the trilinear slepton-Higgs coupling would vary after the
rotation to the mass eigenstate basis and the light Higgs field would 
suffer a wavefunction renormalization.
In addition, if the slepton(s) is nearly massless
then we cannot integrate it out, if we had some non-renormalizable terms would
not be suppressed.

\subsection{Non-Generic Case}

\hspace*{2em} In this section we present a short discussion of the non-generic case
in which there are two light Higgs fields in the final 3D theory.
 If the parameters
are fine tuned we could  have a theory with  two or more light scalar particles whose
interactions are described by some potential \footnote{The authors of reference \cite{carena}
 have suggested a scenario in which the right handed stop and one Higgs are light. For this 
case as they have pointed out one must be  careful with colour (and charge) breaking minima
of the scalar potential, or in the slepton case lepton number violation.}. 
 In this case the infrared behaviour must be studied with
new numerical simulations.

This fine-tuned scenario can be realized in several different ways:\\
- two light Higgses (two doublets, a doublet and a singlet, etc.)\\
- a Higgs and a slepton.\\
- a Higgs and a squark (stop).\\
In this last case the main features are  a screened $SU(3)$ $A_{o}$  field, and   spatial $A_{i}$
gluonic fields which are not decoupled from the squark in the 3D theory. Numerical calculations must also
take this into account and the scalar octet should be
integrated out \footnote{In the generic case for the 
MSSM with one light Higgs, we did not have to worry about the $SU(3)$ fields once the
squarks have been integrated out as they decouple from the rest of the
particles in the theory, even though there is a Debye mass for the longitudinal
gluonic field, etc.}.
For parameters of the MSSM such that two scalars remain light, at the third stage 
 only  $A_{o}$ is integrated out. The expression of 
the two Higgs doublet potential  for the case with two light Higgses is given in appendix B.3.

\section{Other Models}

\hspace*{2em} In appendix C we give the full results of for the parameters of the effective 3D theory for
the 2HDM and the NMSSM. Here we summarize salient characteristics of these models.
 
The reduction of a general
2HDM to a three dimensional effective theory is realized in only two stages and
and the main differences with the MSSM are:\\
- the Higgs couplings are not fixed in terms of the gauge couplings.\\
- there are additional scalar interaction terms.\\
- there are no superpartner contributions to the theory.\\
- the $SU(3)$ gauge bosons completely decouple once the fermions
are integrated out.\\
- the off diagonal Higgs mass term
acquires a dominant contribution on the order of $\sim T^{2}$.\\
All of  the above can in principle
change  considerably the dependence of the critical temperature on the parameters in the theory.

The reduction procedure with the addition  of a singlet superfield to the MSSM  has the following features:\\
- it introduces additional couplings in the scalar and higgsino sector which are not determined
in terms of $g_{s}$ or $g$.\\
- the first stage 3D parameters $G$, $H$, $M_{D}$, $M_{Q_{i}}$, $M_{u_{i}}$, $M_{d_{i}}$
 do not receive additional contributions.\\
- there are additional contributions to the wavefunction renormalizations of $\phi_{1}$ and $\phi_{2}$.\\
- for  values of the parameters for which the mass of the scalar singlet is on the order of the
 $SU(2)$ Debye mass, after the second stage in which squarks and sleptons are integrated out, we are left
with three scalar Higgs fields.

\section{Conclusions}

\hspace*{2em} We have constructed, in the high temperature limit, effective three dimensional theories
for the MSSM, a general 2HDM and the NMSSM  which contain a single light scalar field.
We  obtained the full 1-loop relation between the couplings of the
effective theory  and the underlying 4D couplings and masses. 
For the
case that two  Higgs scalars  are light at the phase transition, we have also given the expression for the two Higgs
doublet  potential  whose infrared behaviour must be studied with
numerical methods. 

The original parameters of these theories
can now be related to physical parameters at the electroweak scale. For the effective
theories containing a single light scalar Higgs, this will allow us to 
evaluate the quantity $x_{c} = \bar{\lambda}_{3}/g_{3}^{2}$ as a function of the physical parameters.
In this way, we can determine for which regions of parameter space the electroweak phase transition may be 
sufficiently first order. The results for the MSSM will be presented elsewhere \cite{ewptmssm}.

\vspace{.2in}
\noindent
{\bf Acknowledgments:} 
I would like to especially thank  Glennys Farrar  for proposing this line
of research, as well as many inspiring discussions  on the physics related to
this paper and comments on the 
manuscript. I  am also indebted to Mikhail Shaposhnikov whom I thank for many useful discussions.
I have benefitted from discussions with P. Arnold and e-mail correspondence with S. Martin.
Research supported in part by Colciencias, Colombia.
\appendix
\newpage

\section{ MSSM in Four Dimensions}

\hspace*{2em} Our 4-dimensional Lagrangian will be a supersymmetric $SU(3)\times SU(2)$ gauge theory
with the same particle content as in the MSSM with the exclusion of $U(1)$
vector particles and corresponding superpartner \cite{haber, kane, dawson}.

The MSSM chiral superfield content is

\begin{equation}
\hat{L} = \left( 
\begin{array}{c}
\hat{\nu_{e}} \\ 
\hat{e}
\end{array}
\right) \hspace{.5in} \hat{E^{c}}
\end{equation}

\begin{equation}
\hat{Q} = \left( 
\begin{array}{c}
\hat{u} \\ 
\hat{d}
\end{array}
\right) \hspace{.5in} \hat{U^{c}} \hspace{.5in} \hat{D^{c}}
\end{equation}

\begin{equation}
\hat{\phi_{1}} = \left( 
\begin{array}{c}
\hat{\phi_{1}^{o}} \\ 
-\hat{\phi_{1}^{-}}
\end{array}
\right) \hspace{.5in} \hat{\phi_{2}} = \left( 
\begin{array}{c}
\hat{\phi_{2}^{+}} \\ 
\hat{\phi_{2}^{o}}
\end{array}
\right)
\end{equation}
When we refer to the scalar component of the superfield we will drop the hat.

Instead of writing out explicitly the full 4D Lagrangian we will
define only the quantities we will need to refer to. In particular, the Yukawa
interactions are derived from the superpotential, which for the MSSM is

\begin{eqnarray}
W & = &\mu(\hat{\phi_{1}^{o}}\hat{\phi_{2}^{o}} + 
\hat{\phi_{1}^{+}}\hat{\phi_{2}^{-}}) +
f_{u}(\hat{\phi_{2}^{o}}\hat{u} - \hat{d}\hat{\phi_{2}^{+}})\hat{U}^{c} \nonumber \\
&+& f_{d}(\hat{\phi_{1}^{o}}\hat{d} + \hat{u}\hat{\phi_{1}^{-}})\hat{D}^{c} \nonumber \\
&+& f_{e}(\hat{\phi_{1}^{o}}\hat{e} + \hat{\nu_{e}}\hat{\phi_{1}^{-}})\hat{E}^{c}.
\end{eqnarray}

In order to maintain supersymmetry's virtue of stabilizing the electroweak
scale via the cancellation of quadratic divergences it is 
standard to introduce SUSY
breaking terms which do not reintroduce this type of divergence, so-called
soft SUSY breaking terms. In particular, there are new scalar interactions
proportional to terms in the superpotential as well as mass terms for
scalars, gauginos and higgsinos. The scalar interactions are obtained
replacing each chiral superfield in the superpotential by its corresponding
scalar component.
Without any further assumptions we would have an extraordinary amount of
parameters which make it extremely difficult to do phenomenology. To
simplify our parameter space we assume above the SUSY breaking scale:\\
 1. A unified gaugino mass, $m_{1/2}, m_{\gluino}$.\\
 2. Common mass for squarks and sleptons $m_{o}^2$.\\
 3. A universal $A$ parameter. In the formulae presented in this paper 
we have kept all Yukawa coupling dependence, although with the
exception of the top Yukawa coupling these contributions generally can be  dropped.

The scalar Higgs self-interactions generate, along
with the soft terms for the scalar Higgs fields, a two Higgs
doublet potential of the form

\begin{eqnarray}
 V(\phi_{1}, \phi_{2})& = & m_{1}^{2} \phi_{1}^{\dagger}\phi_{1} + m_{2}^{2} \phi_{2}^{\dagger}\phi_{2} 
+ m_{3}^{2} (\phi_{1}^{\dagger}\phi_{2} + \phi_{2}^{\dagger}\phi_{1})
+ \lambda_{1} (\phi_{1}^{\dagger}\phi_{1})^{2} \nonumber \\
& + &
\lambda_{2} (\phi_{2}^{\dagger}\phi_{2})^{2} + \lambda_{3}
 (\phi_{1}^{\dagger}\phi_{1})(\phi_{2}^{\dagger}\phi_{2}) +
\lambda_{4} (\phi_{1}^{\dagger}\phi_{2})(\phi_{2}^{\dagger}\phi_{1}) 
\label{potencial}
\end{eqnarray}
in which the quartic couplings are fixed in terms of the
 gauge coupling constants. We comment that in order to
express the scalar potential in this way the $\phi_{1}$ field
has been $SU(2)$ rotated. All $\lambda_{i}$ are real
and fixed by supersymmetry at some high scale to be

\begin{eqnarray}
\lambda_{1} &=& {g^{2}\over 8}, \hspace{2cm} \lambda_{2} = {g^{2}\over 8}\\
\lambda_{3} &=& {g^{2}\over 4}, \hspace{2cm} \lambda_{4} = -{g^{2}\over 2}
\label{lambdasusy}
\end{eqnarray}
in the $g'=0$ limit.
As is well known the model contains
five physical Higgs bosons: a charged pair, two neutral
$CP$-even scalars, and a neutral $CP$-odd scalar  \cite{haber, kane, dawson}.

\section{Explicit Relationships between Parameters}
\subsection{First Stage Parameters}

\hspace*{2em} The explicit relations between the 3D coupling constants and masses
expressed in terms of underlying  4D couplings and the temperature, obtained
as a result of 1-loop integration are given below.  These results reduce
to the  partial results given for the MSSM in the literature 
\cite{zwirner}, as well as the Standard Model results \cite{kajantie},
by taking the appropiate limit.
 The formulae
of Appendix  D  was used to obtain the final results. $N$, $N_{f}$, $N_{s}$
 denote the $SU(N)$ gauge group, number of
fermions doublets and number of scalar doublets, respectively.
 $N_{c}$ is the number of colours and it is taken to be 1 for (s)leptons but we
do not insert an explicit index for simplicity. $N_{sq}$ is the number of squark and
slepton doublets. The index $i$ is a generation index. As the values of the $A$ and $\mu$ parameters
are not known we have kept the explicit dependence on these quantities throughout the calculation.

The thermal masses for the Higgs scalars are given by the evaluation of the
diagrams in figure 1 \footnote{Figures were drawn using feynmf.mf.}:

\begin{eqnarray}
M_{1}^{2} &=&  m_{1}^2(1 + {9\over 4} g^2 {L_{b}\over 16\pi^2} - N_{c}\sum_{i} f_{d_{i}}^2
 {L_{f}\over 16\pi^2}
- {3\over 2}g^2 {L_{f}\over 16\pi^2}) + {3\over 16}g^2 T^2 \nonumber \\
&+& N_{c}\sum_{i} f_{d_{i}}^2 {T^2\over 12}
+ T^2 ({\lambda_{1}\over 2} + {\lambda_{3}\over 6} + {\lambda_{4}\over 12})
+ N_{c}\sum_{i}{f_{d_{i}}^2\over 6}T^2 + N_{sq}(\lambda_{3} + {\lambda_{4}\over 2}){T^2\over 6}
\nonumber \\
& +& {g^2\over 8}T^2 - {L_{b}\over 16\pi^2}(6\lambda_{1}m_{1}^2
+ 2\lambda_{3}m_{2}^{2} + \lambda_{4}m_{2}^{2} + 2N_{c}m_{o}^{2}\sum_{i}f_{d_{i}}^{2} +
N_{c}\sum_{i} (f_{d_{i}}^{2} A^{2} \nonumber \\
& +& f_{u_{i}}^{2} \mu^{2})
+ N_{sq} m_{o}^{2}(2\lambda_{3} + \lambda_{4})
+ 3(\mu^{2}g^{2} + m_{1/2}^{2}g^{2}){L_{f}\over 16\pi^2}
\label{M1}
\end{eqnarray}

\begin{eqnarray}
M_{2}^{2} &= &m_{2}^2(1 + {9\over 4} g^2 {L_{b}\over 16\pi^2} - N_{c}\sum_{i} f_{u_{i}}^2
 {L_{f}\over 16\pi^2}
- {3\over 2}g^2 {L_{f}\over 16\pi^2}) + {3\over 16}g^2 T^2 \nonumber \\
& +& N_{c}\sum_{i} f_{u_{i}}^2 {T^2\over 12}
+  T^2 ({\lambda_{2}\over 2} + {\lambda_{3}\over 6} + {\lambda_{4}\over 12})
+ N_{c}\sum_{i}{f_{u_{i}}^2\over 6}T^2 + N_{sq}(\lambda_{3} + {\lambda_{4}\over 2}){T^2\over 6}
\nonumber \\
& +&
 {g^2\over 8}T^2 - {L_{b}\over 16\pi^2}(6\lambda_{2}m_{2}^2 
+ 2\lambda_{3}m_{1}^{2} + \lambda_{4}m_{1}^{2} + 2N_{c} m_{o}^{2}\sum_{i}f_{u_{i}}^{2} + 
N_{c} \sum_{i}(f_{u_{i}}^{2} A^{2} \nonumber \\
&+& f_{d_{i}}^{2}\mu^{2})
+ N_{sq} m_{o}^{2}(2\lambda_{3} + \lambda_{4})
+ 3(\mu^{2}g^{2} + m_{1/2}^{2}g^{2}){L_{f}\over 16\pi^2}
\label{M2}
\end{eqnarray}

\begin{eqnarray}
M_{3}^{2} & =& m_{3}^2(1 + {9\over 4} g^2 {L_{b}\over 16\pi^2} - N_{c}\sum_{i}({f_{d_{i}}^2\over 2}
+ {f_{u_{i}}^2\over 2})  {L_{f}\over 16\pi^2}- {3\over 2}g^2 {L_{f}\over 16\pi^2}) \nonumber \\
&-& {L_{b}\over 16\pi^2}\left((2\lambda_{4} 
 + \lambda_{3})m_{3}^{2} + N_{c}A\mu\sum_{i}(f_{d_{i}}^{2} + f_{u_{i}}^{2})\right)\nonumber \\
&+& 3(\mu g^{2} m_{1/2}){L_{f}\over 16\pi^2}
\label{M3}
\end{eqnarray}
where

\begin{eqnarray}
L_{b} & = & 2\log {\bar{\mu_{4}} e^{\gamma} \over 4\pi T} \nonumber \\
& = & -\log 4\pi T^{2} + \gamma + \log\mu_{4}^{2} \\
L_{f} &=& L_{b} + 4\ln 2 .
\label{Lb}
\end{eqnarray}
$\mu_{4}$ is the mass scale defined by the $\overline{MS}$ scheme. 
For every 3D parameter, the bracket multiplying the corresponding
4D parameter contains the wavefunction renormalization correction.  We mention specifically that
 the scalar-gauge boson loop contributes only
to the wavefunction renormalization of the field, while the fermionic loops contribute
to the  wavefunction renormalization and the mass.

\begin{figure}
\vskip -190pt
\epsfxsize=6in
\epsfysize=8in
\epsffile{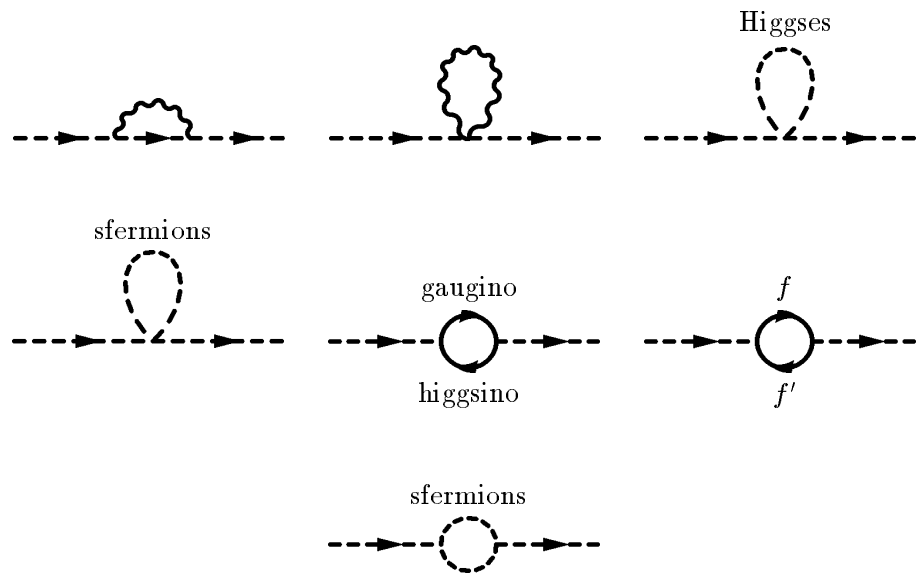}
\vskip -160pt
\caption{Feynman diagrams contributing to the mass of the scalar Higgses and to
wavefunction renormalization.}
\label{mh1figs}
\end{figure}

\begin{figure}
\vskip -180pt
\epsfxsize=6in
\epsfysize=8in
\epsffile{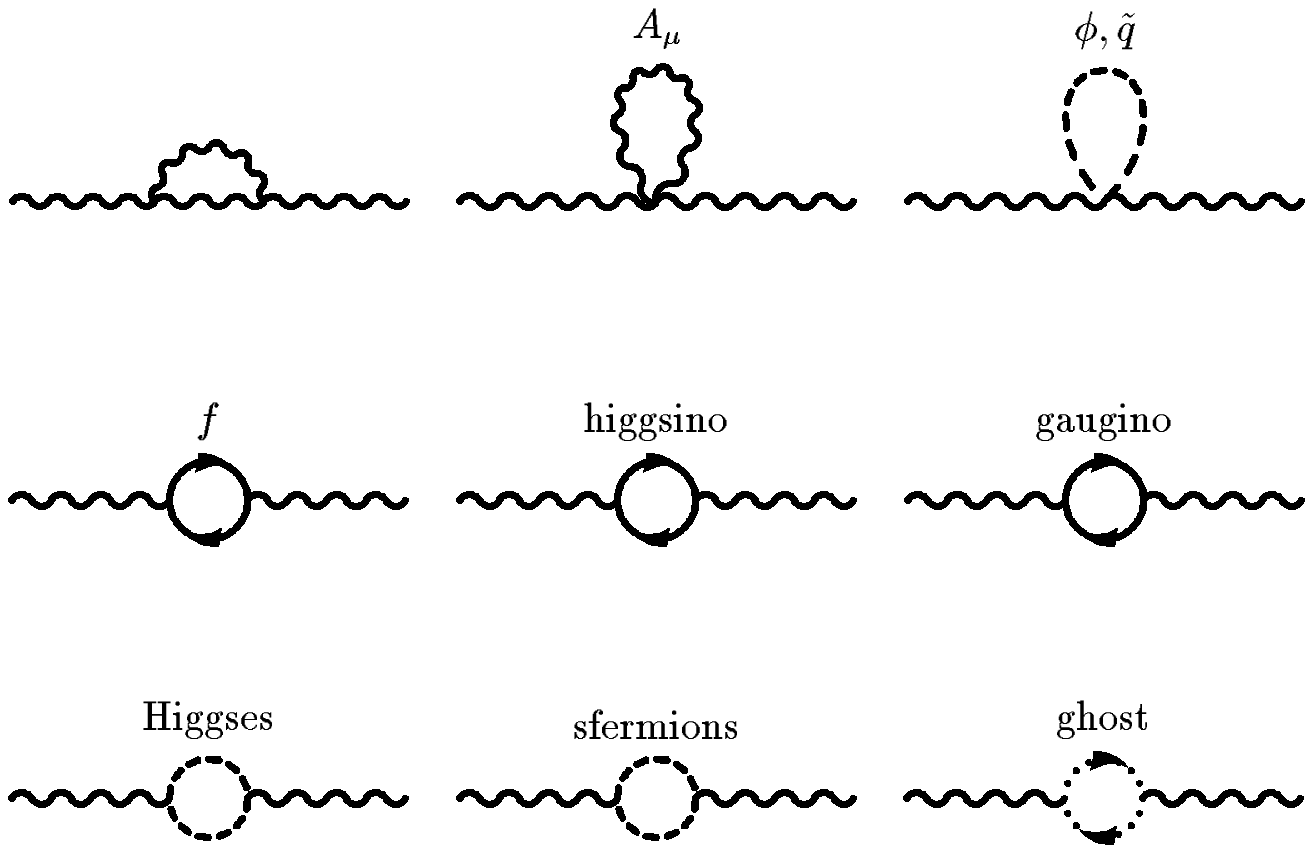}
\vskip -160pt
\caption{Diagrams contributing to the mass   of the $A_{o}$ field and
wavefunction renormalization of the gauge fields. We use a  wavy line for both spatial and temporal
components of the gauge fields.}
\label{mAofigs}
\end{figure}

The Debye mass induced for the temporal component of the  $SU(2)$ gauge field has additional contributions
from those of the Standard Model arising from higgsino, squarks, sleptons and chargino
contributions as shown in figure 2.

\begin{equation}
M_{D}^{2} = {g^2 T^2\over 6}(6 + N_{s} + N_{F}/2 + N_{H}/2).
\label{MD}
\end{equation}

Figure 3 shows the diagrams which contribute to the
 mass terms for a squark (slepton) doublet. For the   up and down right handed squarks (sleptons) 
we must neglect the diagrams with gauge boson and gaugino loops.

\begin{eqnarray}
M_{Q}^{i^{2}} &=& m_{o}^2(1 + {9\over 4} g^2 {L_{b}\over 16\pi^2} - (f_{d_{i}}^2 + f_{u_{i}}^{2})
 {L_{f}\over 16\pi^2}
- {3\over 2}g^2 {L_{f}\over 16\pi^2} + 4g_{s}^{2} {L_{b}\over 16\pi^2} \nonumber \\
&-& {8\over 3}g_{s}^{2} {L_{f}\over 16\pi^2}) + {2\over 3}g_{s}^{2} T^{2} + {3\over 16}g^2 T^2 + (f_{d_{i}}^2 + f_{u_{i}}^{2})
{T^2\over 12}
+ T^2 ({\lambda_{1}\over 2} + {\lambda_{3}\over 3} + {\lambda_{4}\over 6}) \nonumber \\
&+& (f_{d_{i}}^2 + f_{u_{i}}^{2}){T^2\over 6} + {g^2\over 8}T^2 -
 {L_{b}\over 16\pi^2}( {4\over 3} g_{s}^{2} m_{o}^{2} + 6\lambda_{1}m_{o}^2 +
 m_{o}^{2}(f_{u_{i}}^{2} + f_{d_{i}}^{2})
\nonumber \\
& +&
f_{d_{i}}^{2}m_{1}^{2} + f_{u_{i}}^{2}m_{2}^{2} +  (f_{d_{i}}^{2} A^{2} 
+ f_{u_{i}}^{2}\mu^{2} ) + (f_{u_{i}}^{2} A^{2} + f_{d_{i}}^{2}\mu^{2} ) +
2\lambda_{3}(m_{1}^{2} + m_{2}^{2}) \nonumber \\
&+& \lambda_{4}(m_{1}^{2} + m_{2}^{2})) 
+ (3 m_{1/2}^{2}g^{2} + {16\over 3} m_{\gluino}^{2} g_{s}^{2}){L_{f}\over 16\pi^2}
 + 2\mu^{2}(f_{u_{i}}^{2} + f_{d_{i}}^{2})
 {L_{f}\over 16\pi^2} \nonumber \\
&+& {T^{2}\over 24} (N_{sq}-1)(4\lambda_{3} + 2\lambda_{4}) 
- (N_{sq}-1)m_{o}^{2} (2\lambda_{3} + \lambda_{4}){L_{b}\over 16\pi^2}
\label{MoLU}
\end{eqnarray}

\begin{eqnarray}
M_{u}^{i^{2}} &= &m_{o}^{2}(1 - 2f_{u_{i}}^{2}{L_{f}\over 16\pi^2} 
 + 4g_{s}^{2} {L_{b}\over 16\pi^2} - {8\over 3}g_{s}^{2} {L_{f}\over 16\pi^2}) + 
{2\over 3}g_{s}^{2} T^{2} + 
f_{u_{i}}^{2} {T^{2}\over 3} + f_{u_{i}}^{2} {T^{2}\over 6} \nonumber \\
& -& {L_{b}\over 16\pi^2}[{4\over 3} g_{s}^{2} m_{o}^{2} + 
f_{u_{i}}^{2}(2m_{2}^{2} + 2\mu^{2} + 2A^{2} + 2 m_{o}^{2})]
\nonumber \\
&+ &(4\mu^{2}f_{u_{i}}^{2} +  {16\over 3} m_{\gluino}^{2} g_{s}^{2}){L_{f}\over 16\pi^2}
\label{MoRU}
\end{eqnarray}

\begin{figure}
\vskip -180pt
\epsfxsize=6in
\epsfysize=8in
\epsffile{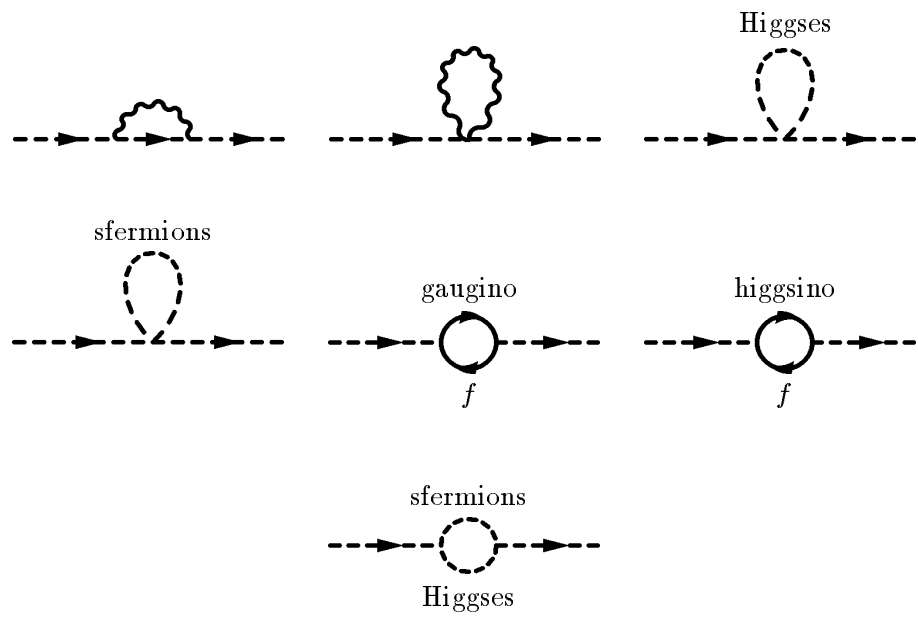}
\vskip -160pt
\caption{Diagrams contributing to the mass and
wavefunction renormalization of the squarks and sleptons.}
\label{squarkmass}
\end{figure}

\begin{eqnarray}
M_{d}^{i^{2}} &=& m_{o}^{2}(1 - 2f_{d_{i}}^{2}{L_{f}\over 16\pi^2}
 + 4g_{s}^{2} {L_{b}\over 16\pi^2} - {8\over 3}g_{s}^{2} {L_{f}\over 16\pi^2}) + 
{2\over 3}g_{s}^{2} T^{2} + 
 f_{d_{i}}^{2}{T^{2}\over 3} +  f_{d_{i}}^{2}{T^{2}\over 6} \nonumber \\
&-& {L_{b}\over 16\pi^2}[{4\over 3} g_{s}^{2} m_{o}^{2} + 
f_{d_{i}}^{2}(2m_{1}^{2} + 2\mu^{2} + 2A^{2} + 2 m_{o}^{2})]
\nonumber \\
&+& (4\mu^{2}f_{d_{i}}^{2} +  {16\over 3} m_{\gluino}^{2} g_{s}^{2}) {L_{f}\over 16\pi^2} ,
\label{MoRD}
\end{eqnarray}
The expressions for the slepton masses are omitted 
although they may be readily obtained by excluding the $g_{s}$ corrections, noting that
there is no right handed sneutrino and dropping all $f_{u_{i}}^{2}$ contributions to
$M_{Q}$, and $M_{d_{i}}$. This is because the sleptons do not have a Yukawa-type coupling to
the $\phi_{2}$ field.

In figure 4 we show the diagrams contributing to the quartic
Higgs couplings.
The full expressions for the scalar couplings
are

\begin{eqnarray}
\Lambda_{1} & =& \lambda_{1}T(1 + {9\over 2} g^2 {L_{b}\over 16\pi^2} - 2 N_{c}\sum_{i} 
f_{d_{i}}^2 {L_{f}\over 16\pi^2}
- 3 g^2 {L_{f}\over 16\pi^2}) + T[-{9\over 16} g^4{L_{b}\over 16\pi^2}\nonumber \\
&+& {3\over 8} {g^4\over 16 \pi^2} + N_{c}\sum_{i} f_{d_{i}}^4 {L_{f}\over 16\pi^2} 
-(12\lambda_{1}^2 + \lambda_{3}^2 + {\lambda_{4}^{2}\over 2} + \lambda_{3}\lambda_{4})
{L_{b}\over 16\pi^2} \nonumber \\
& +& N_{c}[ - (\lambda_{3} + \lambda_{4})\sum_{i}f_{d_{i}}^2 {L_{b}\over 16\pi^2}
- \sum_{i} f_{d_{i}}^4 {L_{b}\over 16\pi^2}]\nonumber \\
&+&N_{sq}[ - {(\lambda_{3} + \lambda_{4})^2\over 2} {L_{b}\over 16\pi^2}
 -{\lambda_{3}^{2}\over 2}]{L_{b}\over 16\pi^2} + {5\over 4} g^4 
{L_{f}\over 16\pi^2}]
\label{lambda1}
\end{eqnarray}

\begin{eqnarray}
\Lambda_{2}& =& \lambda_{2}T(1 + {9\over 2} g^2 {L_{b}\over 16\pi^2} - 2 N_{c}\sum_{i}
 f_{u_{i}}^2 {L_{f}\over 16\pi^2}
- 3 g^2 {L_{f}\over 16\pi^2}) + T[-{9\over 16} g^4{L_{b}\over 16\pi^2} \nonumber \\
&+& {3\over 8} {g^4\over 16 \pi^2} + N_{c}\sum_{i} f_{u_{i}}^4 {L_{f}\over 16\pi^2}
- (12\lambda_{2}^2 + \lambda_{3}^2 + {\lambda_{4}^{2}\over 2} + \lambda_{3}\lambda_{4})
{L_{b}\over 16\pi^2} \nonumber \\
&+& N_{c}[- (\lambda_{3} + \lambda_{4})\sum_{i}f_{u_{i}}^2 {L_{b}\over 16\pi^2}
- \sum_{i} f_{u_{i}}^4 {L_{b}\over 16\pi^2}]\nonumber \\
&+&N_{sq}[ - {(\lambda_{3} +
 \lambda_{4})^2 \over 2}{L_{b}\over 16\pi^2}
- {\lambda_{3}^{2}\over 2}]{L_{b}\over 16\pi^2} + {5\over 4} g^4
 {L_{f}\over 16\pi^2}]
\label{lambda2}
\end{eqnarray}

\begin{eqnarray}
\Lambda_{3} &= &\lambda_{3}T(1 + {9\over 2} g^2 {L_{b}\over 16\pi^2} - N_{c}(f_{u_{i}}^2 +
f_{d_{i}}^2) {L_{f}\over 16\pi^2}
- 3 g^2 {L_{f}\over 16\pi^2}) + T[-{9\over 8} g^4{L_{b}\over 16\pi^2}  \nonumber \\
&+& {3\over 4} {g^4\over 16 \pi^2}
-(6\lambda_{1}\lambda_{3} + 6\lambda_{2}\lambda_{3}+
2\lambda_{1}\lambda_{4}+ 2\lambda_{2}\lambda_{4} + 2\lambda_{3}^2 + \lambda_{4}^{2}
 )
{L_{b}\over 16\pi^2}\nonumber \\
& -& N_{c}\sum_{i}[(\lambda_{3} + \lambda_{4}) (f_{d_{i}}^2 + f_{u_{i}}^2) +
2f_{u_{i}}^{2}f_{d_{i}}^{2}]{L_{b}\over 16\pi^2}
- N_{sq}[(2\lambda_{3}^{2} + 2\lambda_{3} \lambda_{4}\nonumber \\
& +& \lambda_{4}^{2})]
 {L_{b}\over 16\pi^2} 
+ 2N_{c}\sum_{i}f_{u_{i}}^{2}f_{d_{i}}^{2}{L_{f}\over 16\pi^2}
+ {5\over 2} g^4 {L_{f}\over 16\pi^2}]
\label{lambda3}
\end{eqnarray}

\begin{eqnarray}
\Lambda_{4}& =  &\lambda_{4}T(1 + {9\over 2} g^2 {L_{b}\over 16\pi^2} - N_{c}\sum_{i}
(f_{u_{i}}^2 +
f_{d_{i}}^2) {L_{f}\over 16\pi^2}
- 3 g^2 {L_{f}\over 16\pi^2}) + T[-(
2\lambda_{1}\lambda_{4}\nonumber \\
& +& 2\lambda_{2}\lambda_{4} 
+ 2\lambda_{4}^{2}
+ 4\lambda_{3}\lambda_{4})
{L_{b}\over 16\pi^2}+
 N_{c}\sum_{i} [\lambda_{4}(f_{d_{i}}^2 + f_{u_{i}}^2){L_{b}\over 16\pi^2}
 + 2 f_{u_{i}}^{2}f_{d_{i}}^{2}] {L_{b}\over 16\pi^2} \nonumber \\
&+& N_{sq}\lambda_{4}^{2}{L_{b}\over 16\pi^2} - 2N_{c}\sum_{i}
f_{u_{i}}^{2}f_{d_{i}}^{2}{L_{f}\over 16\pi^2}
- 2 g^4 {L_{f}\over 16\pi^2} ].
\label{lambda4}
\end{eqnarray}

\begin{figure}
\vskip -180pt
\epsfxsize=6in
\epsfysize=8in
\epsffile{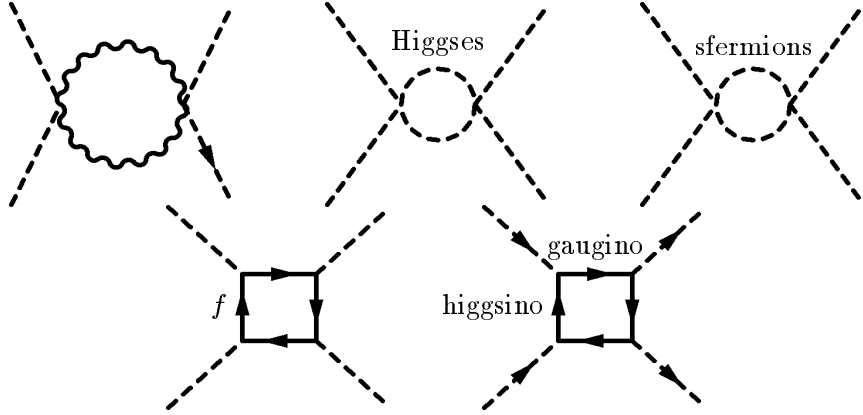}
\vskip -180pt
\caption{Diagrams contributing to the quartic Higgs couplings.}
\label{4higgsfigs}
\end{figure}

There are similar diagrams to those  in figure 4 for the
the quartic couplings of the higgses to squarks and sleptons which have not been shown.
They make the following contributions to these couplings:

\begin{eqnarray}
\bar{f_{d_{i}}^{L}}^{2} &=&  f_{d_{i}}^{2}T( 1 + {9\over 2} g^2 {L_{b}\over 16\pi^2} -
(N_{c}\sum_{i}f_{d_{i}}^2 + f_{d_{i}}^2 +
f_{u_{i}}^2) {L_{f}\over 16\pi^2} - 3 g^2 {L_{f}\over 16\pi^2} \nonumber \\
&+& 4g_{s}^{2} {L_{b}\over 16\pi^2}
- {8\over 3}g_{s}^{2} {L_{f}\over 16\pi^2})
- {T\over 2}[ (\lambda_{4}^{2} 
- \lambda_{4}f_{u_{i}}^{2} + 4\lambda_{4}f_{d_{i}}^{2} + 2f_{d_{i}}^{4}  +
4\lambda_{3}\lambda_{4} \nonumber \\
&+& 4\lambda_{3}f_{d_{i}}^{2} + 2\lambda_{2}\lambda_{4} 
+ 2\lambda_{1}f_{d_{i}}^{2} +
2\lambda_{1}\lambda_{4} + 2\lambda_{2}f_{d_{i}}^{2} + {4\over 3}(\lambda_{4} +
f_{d_{i}}^{2})g_{s}^{2}) {L_{b}\over 16\pi^2} \nonumber \\
&-& [(N_{sq}-1)(\lambda_{4}^{2})
+ N_{c}\lambda_{4}\sum_{i}f_{d_{i}}^{2} 
- \lambda_{4} f_{d_{i}}^{2}]{L_{b}\over 16\pi^2}  \nonumber \\
&+& (2 g^{4} - {16\over 3} f_{d_{i}}^{2}g_{s}^{2}){L_{f}\over 16\pi^2}]
\label{fdL2}
\end{eqnarray}

\begin{eqnarray}
\Lambda_{3}^{Q_{i}1} &=&  \lambda_{3}T( 1  + {9\over 2} g^2 {L_{b}\over 16\pi^2} -
(N_{c}\sum_{i} f_{d_{i}}^2 +  f_{d_{i}}^2 +
f_{u_{i}}^2) {L_{f}\over 16\pi^2} - 3 g^2 {L_{f}\over 16\pi^2} \nonumber \\
&+& 4g_{s}^{2} {L_{b}\over 16\pi^2}
- {8\over 3}g_{s}^{2} {L_{f}\over 16\pi^2})
+ T[ -{9\over 8} g^4{L_{b}\over 16\pi^2} + {3\over 4} {g^4\over 16 \pi^2} 
- ({4\over 3}\lambda_{3} g_{s}^{2} + 4\lambda_{3}^{2} \nonumber \\
&+& 6\lambda_{1}\lambda_{3} + 2\lambda_{4}^{2}
+  2\lambda_{4}f_{d_{i}}^{2} + 2 f_{d_{i}}^{4} + 6\lambda_{2}\lambda_{3} +
2\lambda_{3}\lambda_{4} +  \lambda_{4} f_{u_{i}}^{2}
+ \lambda_{3} f_{u_{i}}^{2} \nonumber \\
&+& 2\lambda_{1}\lambda_{4}
 + 2\lambda_{2}f_{d_{i}}^{2} 
+ 2\lambda_{2}\lambda_{4} + 2\lambda_{1}f_{d_{i}}^{2}) {L_{b}\over 16\pi^2} 
- [(N_{sq}-1)( 2\lambda_{3}^{2} + 2\lambda_{3}\lambda_{4}  \nonumber \\
&+& \lambda_{4}^{2})
+ N_{c}(\lambda_{3} +\lambda_{4})\sum_{i} f_{d_{i}}^{2} -
(\lambda_{3} +\lambda_{4})f_{d_{i}}^{2} ]{L_{b}\over 16\pi^2}\nonumber \\
&+& (2f_{d_{i}}^{4} +  {5 g^{4}\over 2}){L_{f}\over 16\pi^2}]
\label{Lambda3Q1}
\end{eqnarray}

\begin{eqnarray}
\Lambda_{4}^{Q_{i}1} &=&  \lambda_{4}T(1 + {9\over 2} g^2 {L_{b}\over 16\pi^2} -
(N_{c}\sum_{i}f_{d_{i}}^2 + f_{d_{i}}^2 +
f_{u_{i}}^2) {L_{f}\over 16\pi^2} - 3 g^2 {L_{f}\over 16\pi^2} \nonumber \\
&+& 4g_{s}^{2} {L_{b}\over 16\pi^2}
- {8\over 3}g_{s}^{2} {L_{f}\over 16\pi^2})
- {T\over 2}[ (\lambda_{4}^{2} 
- \lambda_{4}f_{u_{i}}^{2} + 4\lambda_{4}f_{d_{i}}^{2} + 2f_{d_{i}}^{4}  +
4\lambda_{3}\lambda_{4} \nonumber \\
&+& 4\lambda_{3}f_{d_{i}}^{2} + 2\lambda_{2}\lambda_{4} 
+ 2\lambda_{1}f_{d_{i}}^{2} +
2\lambda_{1}\lambda_{4} + 2\lambda_{2}f_{d_{i}}^{2} + {4\over 3}(\lambda_{4} +
f_{d_{i}}^{2})g_{s}^{2}) {L_{b}\over 16\pi^2} \nonumber \\
&-& [(N_{sq}-1)(\lambda_{4}^{2})
+ N_{c}\lambda_{4}\sum_{i}f_{d_{i}}^{2} 
- \lambda_{4} f_{d_{i}}^{2}]{L_{b}\over 16\pi^2}  \nonumber \\
&+& (2 g^{4} - {16\over 3} f_{d_{i}}^{2}g_{s}^{2}){L_{f}\over 16\pi^2}]
\label{Lambda4Q1}
\end{eqnarray}

\begin{eqnarray}
\bar{f_{u_{i}}^{L}}^{2} &=&  f_{u_{i}}^{2}T( 1 + {9\over 2} g^2 {L_{b}\over 16\pi^2} -
(N_{c}\sum_{i}f_{u_{i}}^2 + f_{u_{i}}^2 +
f_{d_{i}}^2) {L_{f}\over 16\pi^2} - 3 g^2 {L_{f}\over 16\pi^2} \nonumber \\
&+& 4g_{s}^{2} {L_{b}\over 16\pi^2}
- {8\over 3}g_{s}^{2} {L_{f}\over 16\pi^2})
- {T\over 2}[ (\lambda_{4}^{2} 
- \lambda_{4}f_{d_{i}}^{2} + 4\lambda_{4}f_{u_{i}}^{2} + 2f_{u_{i}}^{4}  +
4\lambda_{3}\lambda_{4}  \nonumber \\
&+& 4\lambda_{3}f_{u_{i}}^{2} + 2\lambda_{2}\lambda_{4} 
+ 2\lambda_{1}f_{u_{i}}^{2} +
2\lambda_{1}\lambda_{4} + 2\lambda_{2}f_{u_{i}}^{2} + {4\over 3}(\lambda_{4} +
f_{u_{i}}^{2})g_{s}^{2}) {L_{b}\over 16\pi^2} \nonumber \\
&-& [(N_{sq}-1)(\lambda_{4}^{2})
+ N_{c}\lambda_{4}\sum_{i}f_{u_{i}}^{2} 
- \lambda_{4} f_{u_{i}}^{2}]{L_{b}\over 16\pi^2} \nonumber \\
&+& (2 g^{4} - {16\over 3} f_{u_{i}}^{2}g_{s}^{2}){L_{f}\over 16\pi^2}]
\label{fuL2}
\end{eqnarray}

\begin{eqnarray}
\Lambda_{3}^{Q_{i}2} &=&  \lambda_{3}T( 1  + {9\over 2} g^2 {L_{b}\over 16\pi^2} -
(N_{c}\sum_{i} f_{u_{i}}^2 +  f_{u_{i}}^2 +
f_{d_{i}}^2) {L_{f}\over 16\pi^2} - 3 g^2 {L_{f}\over 16\pi^2}  \nonumber \\
&+& 4g_{s}^{2} {L_{b}\over 16\pi^2}
- {8\over 3}g_{s}^{2} {L_{f}\over 16\pi^2})
+ T[ -{9\over 8} g^4{L_{b}\over 16\pi^2} + {3\over 4} {g^4\over 16 \pi^2} 
- ({4\over 3}\lambda_{3} g_{s}^{2} + 4\lambda_{3}^{2}  \nonumber \\
&+& 6\lambda_{1}\lambda_{3} + 2\lambda_{4}^{2}
+  2\lambda_{4}f_{u_{i}}^{2} + 2 f_{u_{i}}^{4} + 6\lambda_{2}\lambda_{3} +
2\lambda_{3}\lambda_{4} +  \lambda_{4} f_{d_{i}}^{2}
+ \lambda_{3} f_{d_{i}}^{2}  \nonumber \\
&+& 2\lambda_{1}\lambda_{4}
 + 2\lambda_{2}f_{u_{i}}^{2}
+ 2\lambda_{2}\lambda_{4} + 2\lambda_{1}f_{u_{i}}^{2}) {L_{b}\over 16\pi^2} 
- [(N_{sq}-1)( 2\lambda_{3}^{2} + 2\lambda_{3}\lambda_{4} \nonumber \\
&+& \lambda_{4}^{2}) 
 N_{c}(\lambda_{3} +\lambda_{4})\sum_{i} f_{u_{i}}^{2} -
(\lambda_{3} +\lambda_{4})f_{u_{i}}^{2} ]{L_{b}\over 16\pi^2}\nonumber \\
&+& (2f_{u_{i}}^{4} +  {5 g^{4}\over 2}){L_{f}\over 16\pi^2}]
\label{Lambda3Q2}
\end{eqnarray}

\begin{eqnarray}
\Lambda_{4}^{Q_{i}2} &=&  \lambda_{4}T( 1 + {9\over 2} g^2 {L_{b}\over 16\pi^2} -
(N_{c}\sum_{i}f_{u_{i}}^2 + f_{u_{i}}^2 +
f_{d_{i}}^2) {L_{f}\over 16\pi^2} - 3 g^2 {L_{f}\over 16\pi^2} \nonumber \\
&+& 4g_{s}^{2} {L_{b}\over 16\pi^2}
- {8\over 3}g_{s}^{2} {L_{f}\over 16\pi^2})
- {T\over 2}[ (\lambda_{4}^{2} 
- \lambda_{4}f_{d_{i}}^{2} + 4\lambda_{4}f_{u_{i}}^{2} + 2f_{u_{i}}^{4}  +
4\lambda_{3}\lambda_{4}  \nonumber \\
&+& 4\lambda_{3}f_{u_{i}}^{2} + 2\lambda_{2}\lambda_{4} 
+ 2\lambda_{1}f_{u_{i}}^{2} +
2\lambda_{1}\lambda_{4} + 2\lambda_{2}f_{u_{i}}^{2} + {4\over 3}(\lambda_{4} +
f_{u_{i}}^{2})g_{s}^{2}) {L_{b}\over 16\pi^2} \nonumber \\
&-& [(N_{sq}-1)(\lambda_{4}^{2})
+ N_{c}\lambda_{4}\sum_{i}f_{u_{i}}^{2} 
- \lambda_{4} f_{u_{i}}^{2}]{L_{b}\over 16\pi^2} \nonumber \\
&+& (2 g^{4} - {16\over 3} f_{u_{i}}^{2}g_{s}^{2}){L_{f}\over 16\pi^2}]
\label{Lambda4Q2}
\end{eqnarray}

\begin{eqnarray}
\bar{f_{d_{i}}^{R}}^{2} &= &f_{d_{i}}^{2}T(1 + {9\over 4}g^{2}{L_{b}\over 16\pi^2} - 
( N_{c}\sum_{i} f_{d_{i}}^2 + 2f_{d_{i}}^2 ) {L_{f}\over 16\pi^2} -
 {3\over 2} g^2 {L_{f}\over 16\pi^2} \nonumber \\
&+& 4g_{s}^{2} {L_{b}\over 16\pi^2} - {8\over 3}g_{s}^{2} {L_{f}\over 16\pi^2})
+ T[-{L_{b}\over 16\pi^2}( {4\over 3} f_{d_{i}}^{2} g_{s}^{2} + 2\lambda_{3} + \lambda_{4} \nonumber \\
&+& f_{d_{i}}^{2} f_{u_{i}}^{2} + 3f_{d_{i}}^{4} +6\lambda_{1}f_{d_{i}}^{2}) + 
{L_{f}\over 16\pi^2}({16\over 3} f_{d_{i}}^{2} g_{s}^{2} + 2f_{d_{i}}^{4} + 3f_{d_{i}}^{2} g^{2})]
\label{fdR2}
\end{eqnarray}

\begin{eqnarray}
\bar{f_{u_{i}}^{R}}^{2} &=& f_{u_{i}}^{2}T(1 + {9\over 4}g^{2}{L_{b}\over 16\pi^2} - 
( N_{c}\sum_{i} f_{u_{i}}^2 + 2  f_{u_{i}}^2)
 {L_{f}\over 16\pi^2} - {3\over 2} g^2 {L_{f}\over 16\pi^2}) \nonumber \\
&+& 4g_{s}^{2} {L_{b}\over 16\pi^2} - {8\over 3}g_{s}^{2} {L_{f}\over 16\pi^2})
+ T[-{L_{b}\over 16\pi^2}({4\over 3} f_{u_{i}}^{2} g_{s}^{2} + 2\lambda_{3} + \lambda_{4} \nonumber \\
&+& f_{d_{i}}^{2} f_{u_{i}}^{2} +3f_{u_{i}}^{4} +6\lambda_{2}f_{u_{i}}^{2}) + 
{L_{f}\over 16\pi^2}({16\over 3} f_{u_{i}}^{2} g_{s}^{2} + 2f_{u_{i}}^{4} + 3f_{u_{i}}^{2} g^{2})]
\label{fuR2}
\end{eqnarray}
We would like to point out that the ${T\over 2}$ factor in the 
expressions for $\Lambda_{4}^{Q_{i}1}$, $\bar{f_{d_{i}}^{L}}^{2}$, $\Lambda_{4}^{Q_{i}2}$
and $\bar{f_{u_{i}}^{L}}^{2}$ is due to the fact that  the sum of the first two
is the full quartic coupling to the $\phi_{1}$ field, and analogously for the second pair.

\begin{figure}
\vskip -180pt
\epsfxsize=6in
\epsfysize=8in
\epsffile{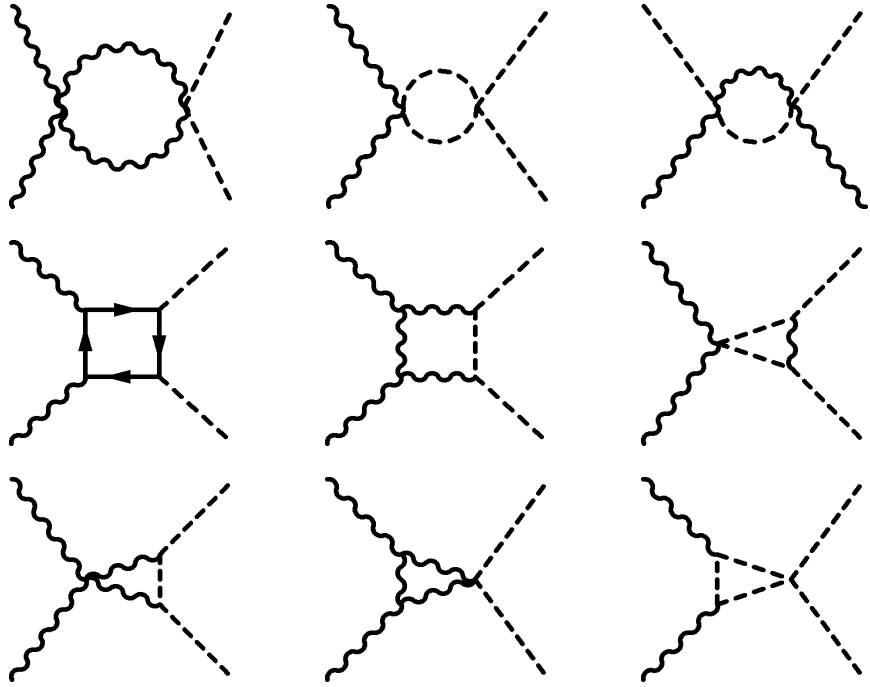}
\vskip -160pt
\caption{Diagrams contributing to the gauge and quartic $A_{o}$-scalar couplings.}  
\label{gaugecoupfigs}
\end{figure}

 As explained in \cite{kajantie} we can obtain the gauge and quartic $A_{o}$-scalar doublet
couplings from the same set of diagrams depicted in figure 5:

\begin{eqnarray}
G^{2} &=& g^{2}T[1 + {g^{2}\over 16\pi^{2}}({44- N_{s}\over 6}L_{b} - {1\over 3}(N_{f}
+ N_{H} + 4)L_{f} + {2\over 3})]
\label{G2}
\end{eqnarray}

\begin{eqnarray}
H &=& {g^{2}T\over 4}[1 + {g^{2}\over 16\pi^{2}}({44- N_{s}\over 6}L_{b} - {1\over 3}(N_{f}
+  N_{H} + 4)L_{f}) \nonumber \\
& +& {1\over 16\pi^{2}}({35\over 6}g^{2} - {N_{s}\over 3} g^{2} + {g^{2}\over 3}
(N_{F} + N_{H} + 4) + 12 \lambda_{1}\nonumber \\
& + &2(N_{s}-1)(2\lambda_{3}+\lambda_{4}))].
\label{H}
\end{eqnarray}

The scalar trilinear couplings are also modified as can be seen in figure 6:

\begin{eqnarray}
\bar{A}f_{u_{i}} & = & Af_{u_{i}}T^{1/2}(1 + {9\over 4} g^2 {L_{b}\over 16\pi^2} - 
({f_{d_{i}}^2\over 2} +  {N_{c}\over 2}\sum_{i}f_{u_{i}}^{2} + {3\over 2}f_{u_{i}}^{2}) 
{L_{f}\over 16\pi^2}
- {3\over 2}g^2 {L_{f}\over 16\pi^2} \nonumber \\
&+& 4g_{s}^{2} {L_{b}\over 16\pi^2} - {8\over 3}g_{s}^{2} {L_{f}\over 16\pi^2})
- Af_{u_{i}}T^{1/2}(\lambda_{3} + 2\lambda_{4} + 3f_{u_{i}}^{2}
+ N_{c} \sum_{i}f_{u_{i}}^{2} + f_{d_{i}}^{2} \nonumber \\
& -& {4\over 3} g_{s}^{2} ){L_{b}\over 16\pi^2}
- m_{1/2} f_{u_{i}}T^{1/2}(3g^{2}){L_{f}\over 16\pi^2} -  m_{\gluino} 
f_{u_{i}}T^{1/2}({16\over 3}g_{s}^{2}){L_{f}\over 16\pi^2}
\label{Afu}
\end{eqnarray}

\begin{eqnarray}
\bar{A}f_{d_{i}} & = & Af_{d_{i}}T^{1/2}(1 + {9\over 4} g^2 {L_{b}\over 16\pi^2} - 
({f_{u_{i}}^2\over 2} +  {N_{c}\over 2}\sum_{i}f_{d_{i}}^{2} + {3\over 2}f_{d_{i}}^{2}) 
{L_{f}\over 16\pi^2}
- {3\over 2}g^2 {L_{f}\over 16\pi^2} \nonumber \\
&+& 4g_{s}^{2} {L_{b}\over 16\pi^2} - {8\over 3}g_{s}^{2} {L_{f}\over 16\pi^2})
- Af_{d_{i}}T^{1/2}(\lambda_{3} + 2\lambda_{4} + 3f_{d_{i}}^{2}
+ N_{c} \sum_{i}f_{d_{i}}^{2} + f_{u_{i}}^{2} \nonumber \\
& -&{4\over 3} g_{s}^{2} ){L_{b}\over 16\pi^2}
- m_{1/2} f_{d_{i}}T^{1/2}(3g^{2}){L_{f}\over 16\pi^2} -  m_{\gluino} 
f_{d_{i}}T^{1/2}({16\over 3}g_{s}^{2}){L_{f}\over 16\pi^2}
\label{Afd}
\end{eqnarray}

\begin{figure}
\vskip -180pt
\epsfxsize=6in
\epsfysize=6in
\epsffile{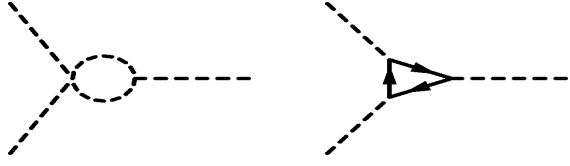}
\vskip -150pt
\caption{Diagrams contributing to the trilinear scalar couplings.}  
\label{trilinear}
\end{figure}

\begin{eqnarray}
\bar{\mu}f_{u_{i}} & = & \mu f_{u_{i}}T^{1/2}(1 +  {9\over 4} g^2 {L_{b}\over 16\pi^2}
 - ({N_{c}\over 2}\sum_{i} f_{d_{i}}^2 + {1\over 2} f_{d_{i}}^2 +
{3\over 2}f_{u_{i}}^{2}) {L_{f}\over 16\pi^2}
- {3\over 2}g^2 {L_{f}\over 16\pi^2} \nonumber \\
&+& 4g_{s}^{2} {L_{b}\over 16\pi^2} - {8\over 3}g_{s}^{2} {L_{f}\over 16\pi^2})
- \mu f_{u_{i}}T^{1/2}(\lambda_{3} - \lambda_{4} +
N_{c}\sum_{i}f_{u_{i}}^{2} -2 f_{d_{i}}^{2} \nonumber \\
&-&{4\over 3} g_{s}^{2} ){L_{b}\over 16\pi^2}
+ \mu f_{u_{i}}T^{1/2}(2f_{d_{i}}^{2}-3g^{2}){L_{f}\over 16\pi^2}
\label{mufu}
\end{eqnarray}

\begin{eqnarray}
\bar{\mu}f_{d_{i}} & = & \mu f_{d_{i}}T^{1/2}(1 +  {9\over 4} g^2 {L_{b}\over 16\pi^2}
 - ({N_{c}\over 2}\sum_{i}f_{u_{i}}^2 + {1\over 2} f_{u_{i}}^2 +
{3\over 2}f_{d_{i}}^{2}) {L_{f}\over 16\pi^2}
- {3\over 2}g^2 {L_{f}\over 16\pi^2} \nonumber \\
&+& 4g_{s}^{2} {L_{b}\over 16\pi^2} - {8\over 3}g_{s}^{2} {L_{f}\over 16\pi^2})
- \mu f_{d_{i}}T^{1/2}(\lambda_{3} - \lambda_{4} +
N_{c}\sum_{i}f_{d_{i}}^{2} -2 f_{u_{i}}^{2} \nonumber \\
&-& {4\over 3} g_{s}^{2} ){L_{b}\over 16\pi^2}
+ \mu f_{d_{i}}T^{1/2}(2f_{u_{i}}^{2}-3g^{2}){L_{f}\over 16\pi^2}.
\label{mufd}
\end{eqnarray}
We comment that at zero temperature the beta function coefficients 
for the four trilinear couplings given above, which are products of two
parameters, can be obtained from the running of each parameter separately. 
This is true up to an arbitrary
number of loops.

\subsection{Second Stage Parameters}

\hspace*{2em} The gauge coupling is given by the expression

\begin{equation}
\bar{G}^{2} = G^{2}(1- \sum_{i}N_{c}{G^{2}\over 24\pi M_{Q}^{i}}).
\label{Gbar2}
\end{equation}
A clear difference appears now  in the coupling of $A_{o}$ to $\phi_{1}$
and $\phi_{2}$ which is not protected by any symmetry. How large  this difference is 
depends strongly
on the values of the soft-breaking parameters. In general, the expressions
simplify considerably if we ignore the trilinear scalar interaction
terms. Additional diagrams,  shown in figure 7,
which were suppressed by powers of $T^{-2}$ for the
first stage are
included. We  point out that the box diagram
with two external scalar Higgs legs only contributes to the four point function which determines
$\bar{G}^{2}$, as there is no trilinear $A_{o}\phi\phi$ interactions in the
three dimensional theory.

\begin{eqnarray}
\bar{H_{1}} &= & H(1 - \sum_{i}{N_{c}f_{d_{i}}^{2}\bar{A}^{2}\over 12\pi(M_{Q}^{i} +
M_{d}^{i})^{3}} - \sum_{i}{N_{c}f_{u_{i}}^{2}\bar{\mu}^{2}\over 12\pi(M_{Q}^{i} +
M_{u}^{i})^{3}}) \nonumber \\
&-&\sum_{i} N_{c}[ H(2\Lambda_{3}^{Q_{i}1} + \Lambda_{4}^{Q_{i}1} )
{1\over 8\pi M_{Q}^{i}} +  H \bar{f_{d_{i}}^{L}}^{2} {1\over 8\pi M_{Q}^{i}}
\nonumber \\
&-& H{1\over 8\pi M_{Q}^{i}}({f_{d_{i}}^{2}\bar{A}^{2}
\over (M_{Q}^{i} + M_{d}^{i})^{2}} 
+ {f_{u_{i}}^{2}\bar{\mu}^{2}\over(M_{Q}^{i} + M_{u}^{i})^{2}})]
\label{H1}
\end{eqnarray}

\begin{eqnarray}
\bar{H_{2}} &= & H(1 - \sum_{i}{N_{c}f_{u_{i}}^{2}\bar{A}^{2}\over 12\pi(M_{Q}^{i} +
M_{u}^{i})^{3}} - \sum_{i} {N_{c}f_{d_{i}}^{2}\bar{\mu}^{2}\over 12\pi(M_{Q}^{i} +
M_{d}^{i})^{3}}) \nonumber \\
&-& \sum_{i} N_{c}[ H(2\Lambda_{3}^{Q_{i}2} + \Lambda_{4}^{Q_{i}2} )
{1\over 8\pi M_{Q}^{i}} + H\bar{f_{u_{i}}^{L}}^{2}{1\over 8\pi M_{Q}^{i}}\nonumber \\
&-& H{1\over 8\pi M_{Q}^{i}}({f_{u_{i}}^{2}\bar{A}^{2}
\over (M_{Q}^{i} + M_{u}^{i})^{2}}
+ {f_{d_{i}}^{2}\bar{\mu}^{2}\over
(M_{Q}^{i} + M_{d}^{i})^{2}})].
\label{H2}
\end{eqnarray}

\begin{figure}
\vskip -180pt
\epsfxsize=6in
\epsfysize=8in
\epsffile{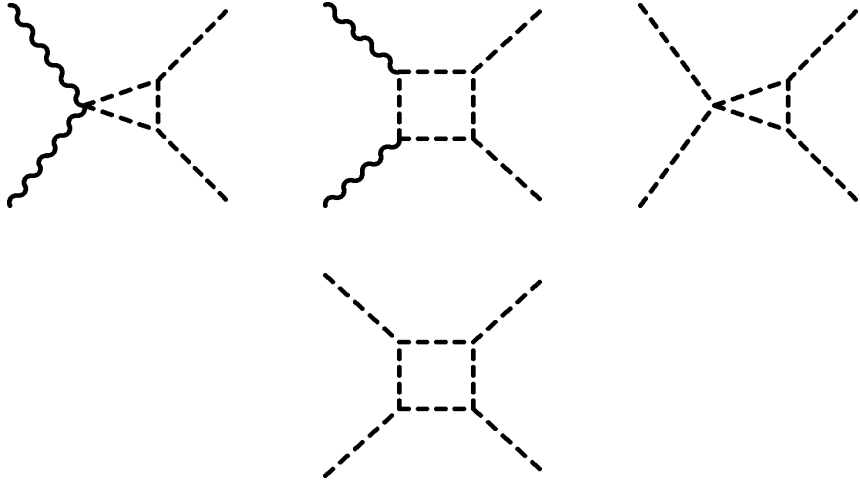}
\vskip -180pt
\caption{Additional diagrams included in the second stage.}  
\label{additfigs}
\end{figure}

The elements of the scalar Higgs doublet mass matrix are now given by:

\begin{eqnarray}
\bar{M_{1}}^{2} & = & M_{1}^{2}(1 - \sum_{i}{N_{c}f_{d_{i}}^{2}\bar{A}^{2}\over 12\pi(M_{Q}^{i} +
M_{d}^{i})^{3}} - \sum_{i}{N_{c}f_{u_{i}}^{2}\bar{\mu}^{2}\over 12\pi(M_{Q}^{i} +
M_{u}^{i})^{3}}) \nonumber \\
&-& \sum_{i} N_{c}[ (\bar{f_{d_{i}}^{L}}^{2}+ 2\Lambda_{3}^{Q_{i}1} +\Lambda_{4}^{Q_{i}1}) {M_{Q}^{i}\over 4\pi} 
+  \bar{f_{d_{i}}^{R}}^{2} {M_{d}^{i}\over 4\pi} \nonumber \\
&+&  {f_{d_{i}}^{2}\bar{A}^{2}\over 
4\pi(M_{Q}^{i} + M_{d}^{i})} + {f_{u_{i}}^{2}\bar{\mu}^{2}\over 4\pi(M_{Q}^{i} +
M_{u}^{i})}]
\label{M1bar2}
\end{eqnarray}

\begin{eqnarray}
\bar{M_{2}}^{2} & = & M_{2}^{2}(1 - \sum_{i}{N_{c}f_{u_{i}}^{2}\bar{A}^{2}\over 12\pi(M_{Q}^{i} +
M_{u}^{i})^{3}} - \sum_{i}{N_{c}f_{d_{i}}^{2}\bar{\mu}^{2}\over 12\pi(M_{Q}^{i} +
M_{d}^{i})^{3}}) \nonumber \\
&-& \sum_{i} N_{c}[ (\bar{f_{u_{i}}^{L}}^{2} + 2\Lambda_{3}^{Q_{i}2} + \Lambda_{4}^{Q_{i}2})  {M_{Q}^{i}\over 4\pi} 
+ \bar{f_{u_{i}}^{R}}^{2} {M_{u}^{i}\over 4\pi} \nonumber \\
& + & {f_{u_{i}}^{2}\bar{A}^{2}\over 
4\pi(M_{Q}^{i} + M_{u}^{i})} + {f_{d_{i}}^{2}\bar{\mu}^{2}\over 4\pi(M_{Q}^{i} +
M_{d}^{i})}]
\label{M2bar2}
\end{eqnarray}

\begin{eqnarray}
\bar{M_{3}}^{2} & = & M_{3}^{2}(1 - \sum_{i}{N_{c}f_{d_{i}}^{2}\bar{A}^{2}\over 24\pi(M_{Q}^{i} +
M_{d}^{i})^{3}} - \sum_{i} {N_{c}f_{u_{i}}^{2}\bar{\mu}^{2}\over 24\pi(M_{Q}^{i} +
M_{u}^{i})^{3}}  \nonumber \\
&-&  \sum_{i}{N_{c}f_{u_{i}}^{2}\bar{A}^{2}\over 24\pi(M_{Q}^{i} +
M_{u}^{i})^{3}}
-\sum_{i}{N_{c}f_{d_{i}}^{2}\bar{\mu}^{2}\over 24\pi(M_{Q}^{i} +
M_{d}^{i})^{3}})  \nonumber \\
&-& \sum_{i}N_{c}[{f_{u_{i}}^{2}\bar{A}\bar{\mu}\over 
4\pi(M_{Q}^{i} + M_{u}^{i})} + {f_{d_{i}}^{2}\bar{\mu}\bar{A}\over 4\pi(M_{Q}^{i} +
M_{d}^{i})}] .
\label{M3bar3}
\end{eqnarray}
The resulting Debye mass from the second stage of integration is

\begin{equation}
\bar{M_{D}}^{2} = M_{D}^{2} - \sum_{i} N_{c}H{M_{Q}^{i}\over 4\pi}.
\label{MDbar}
\end{equation}
We remark that the previous  quantity is always positive for values of the
parameters for which both perturbation theory and the high
temperature expansion are valid.

The resulting quartic Higgs couplings are:

\begin{eqnarray}
\bar{\Lambda}_{1} & = & \Lambda_{1}( 1 - \sum_{i} {N_{c}f_{d_{i}}^{2}\bar{A}^{2}\over 6\pi(M_{Q}^{i} +
M_{d}^{i})^{3}} - \sum_{i}{N_{c}f_{u_{i}}^{2}\bar{\mu}^{2}\over 6\pi(M_{Q}^{i} +
M_{u}^{i})^{3}}) \nonumber \\ 
&+& \sum_{i} N_{c}[- {(\Lambda_{3}^{Q_{i}1} + \Lambda_{4}^{Q_{i}1})^{2}\over 2}{1\over 8\pi M_{Q}^{i}}
- {\Lambda_{3}^{Q1^{2}}\over 2}{1\over 8\pi M_{Q}^{i}} \nonumber \\
&-& (\Lambda_{3}^{Q_{i}1} + \Lambda_{4}^{Q_{i}1})\bar{f_{d_{i}}^{L}}^{2} {1\over 8\pi M_{Q}^{i}} 
- {\bar{f_{d_{i}}^{L}}^{4}\over 2}{1\over 8\pi M_{Q}^{i}}
- {\bar{f_{d_{i}}^{R}}^{4}\over 2}{1\over 8\pi M_{d}^{i}} \nonumber \\
&+& f_{d_{i}}^{2}\bar{A}^{2}
(\Lambda_{3}^{Q_{i}1} + \Lambda_{4}^{Q_{i}1} + \bar{f_{d_{i}}^{L}}^{2}){1\over 8\pi M_{Q}^{i}}
{1\over (M_{Q}^{i} + M_{d}^{i})^{2}}  \nonumber \\
&+& f_{u_{i}}^{2}\bar{\mu}^{2}\Lambda_{3}^{Q_{i}1}
{1\over 8\pi M_{Q}^{i}}{1\over (M_{Q}^{i} + M_{u}^{i})^{2}}
+ f_{d_{i}}^{2}\bar{A}^{2}\bar{f_{d_{i}}^{R}}^{2}
{1\over 8\pi M_{d}^{i}}{1\over (M_{Q}^{i} + M_{d}^{i})^{2}}\nonumber \\
&-& f_{u_{i}}^{4}\bar{\mu}^{4}{1\over 8\pi  M_{u}^{i} M_{Q}^{i}}{1\over (M_{Q}^{i} + M_{u}^{i})^{3}}
-  f_{d_{i}}^{4}\bar{A}^{4} {1\over 8\pi M_{d}^{i} M_{Q}^{i}}{1\over (M_{Q}^{i} + M_{d}^{i})^{3}} ]
\label{Lambda1bar}
\end{eqnarray}

\begin{eqnarray}
\bar{\Lambda}_{2} & = & \Lambda_{2}( 1 - \sum_{i} {N_{c}f_{u_{i}}^{2}\bar{A}^{2}\over 6\pi(M_{Q}^{i} +
M_{u}^{i})^{3}} - \sum_{i}{N_{c}f_{d_{i}}^{2}\bar{\mu}^{2}\over 6\pi(M_{Q}^{i} +
M_{d}^{i})^{3}}) \nonumber \\ 
&+& \sum_{i} N_{c}[-{(\Lambda_{3}^{Q_{i}2} + \Lambda_{4}^{Q_{i}2})^{2}\over 2}{1\over 8\pi M_{Q}^{i}}
- {\Lambda_{3}^{Q2^{2}}\over 2}{1\over 8\pi M_{Q}^{i}} \nonumber \\
&-& (\Lambda_{3}^{Q_{i}2} + \Lambda_{4}^{Q_{i}2})\bar{f_{u_{i}}^{L}}^{2} {1\over 8\pi M_{Q}^{i}}
- {\bar{f_{u_{i}}^{L}}^{4}\over 2}{1\over 8\pi M_{Q}^{i}} 
- {\bar{f_{u_{i}}^{R}}^{4}\over 2}{1\over 8\pi M_{u}^{i}} \nonumber \\
&+& f_{u_{i}}^{2}\bar{A}^{2}
(\Lambda_{3}^{Q_{i}2} + \Lambda_{4}^{Q_{i}2} + \bar{f_{u_{i}}^{L}}^{2}){1\over 8\pi M_{Q}^{i}}
{1\over (M_{Q}^{i} + M_{u}^{i})^{2}}\nonumber \\
&+& f_{d_{i}}^{2}\bar{\mu}^{2}\Lambda_{3}^{Q_{i}2}
{1\over 8\pi M_{Q}^{i}}{1\over (M_{Q}^{i} + M_{d}^{i})^{2}} 
+ f_{u_{i}}^{2}\bar{A}^{2}\bar{f_{u_{i}}^{R}}^{2}
{1\over 8\pi M_{u}^{i}}{1\over (M_{Q}^{i} + M_{u}^{i})^{2}} \nonumber \\
&-& f_{d_{i}}^{4}\bar{\mu}^{4}{1\over 8\pi  M_{d}^{i} M_{Q}^{i}}{1\over (M_{Q}^{i} + M_{d}^{i})^{3}}
-  f_{u_{i}}^{4}\bar{A}^{4} {1\over 8\pi M_{u}^{i} M_{Q}^{i}}{1\over (M_{Q}^{i} + M_{u}^{i})^{3}}]
\label{Lambda2bar}
\end{eqnarray}

\begin{eqnarray}
\bar{\Lambda}_{3} & =& \Lambda_{3}(1 - \sum_{i}{N_{c}f_{d_{i}}^{2}\bar{A}^{2}\over 12\pi(M_{Q}^{i} +
M_{d}^{i})^{3}} - \sum_{i}{N_{c}f_{u_{i}}^{2}\bar{\mu}^{2}\over 12\pi(M_{Q}^{i} +
M_{u}^{i})^{3}}  \nonumber \\
 &-& \sum_{i} {N_{c}f_{u_{i}}^{2}\bar{A}^{2}\over 12\pi(M_{Q}^{i} +
M_{u}^{i})^{3}} 
- \sum_{i}{N_{c}f_{d_{i}}^{2}\bar{\mu}^{2}\over 12\pi(M_{Q}^{i} +
M_{d}^{i})^{3}}) \nonumber \\
&+& \sum_{i} N_{c}[ -\left( 2\Lambda_{3}^{Q_{i}1}\Lambda_{3}^{Q_{i}2}+
\Lambda_{4}^{Q_{i}1}\Lambda_{4}^{Q_{i}2} + \Lambda_{3}^{Q_{i}1} \Lambda_{4}^{Q_{i}2}
+  \Lambda_{3}^{Q_{i}2} \Lambda_{4}^{Q_{i}1}\right){1\over 8\pi M_{Q}^{i}}\nonumber \\
&-& (\bar{f_{u_{i}}^{L}}^{2} \Lambda_{4}^{Q_{i}1} + 
\bar{f_{d_{i}}^{L}}^{2} \Lambda_{4}^{Q_{i}2}
+ 2 \bar{f_{u_{i}}^{L}}^{2}\bar{f_{d_{i}}^{L}}^{2}+
\bar{f_{u_{i}}^{L}}^{2} \Lambda_{3}^{Q_{i}1} 
+ \bar{f_{d_{i}}^{L}}^{2} \Lambda_{3}^{Q_{i}2} ){1\over 8\pi M_{Q}^{i}}\nonumber \\
&+& f_{u_{i}}^{2}(\bar{f_{d_{i}}^{L}}^{2}\bar{A}^{2} + \bar{\mu}^{2}\Lambda_{3}^{Q_{i}2}
+ \bar{A}^{2}\Lambda_{4}^{Q_{i}1} + \Lambda_{3}^{Q_{i}1}\bar{A}^{2})
{1\over 8\pi M_{Q}^{i}}{1\over (M_{Q}^{i} + M_{u}^{i})^{2}} \nonumber \\
&+&  f_{d_{i}}^{2}(\bar{f_{u_{i}}^{L}}^{2}\bar{A}^{2} + \bar{\mu}^{2}\Lambda_{3}^{Q_{i}1}
+ \bar{A}^{2}\Lambda_{4}^{Q_{i}2} + \Lambda_{3}^{Q_{i}2}\bar{A}^{2})
{1\over 8\pi M_{Q}^{i}}{1\over (M_{Q}^{i} + M_{d}^{i})^{2}} \nonumber \\
&+& f_{u_{i}}^{2}\bar{\mu}^{2}\bar{f_{u_{i}}^{R}}^{2}
{1\over 8\pi M_{u}^{i}}{1\over (M_{Q}^{i} + M_{u}^{i})^{2}}
+ f_{d_{i}}^{2}\bar{\mu}^{2}\bar{f_{d_{i}}^{R}}^{2}
{1\over 8\pi M_{d}^{i}}{1\over (M_{Q}^{i} + M_{d}^{i})^{2}}\nonumber \\
&-& 2( (f_{d_{i}}^{2} f_{u_{i}}^{2}\bar{\mu}^{4} 
- 2  f_{d_{i}}^{2} f_{u_{i}}^{2}\bar{A}^{2}\bar{\mu}^{2} 
+  f_{d_{i}}^{2} f_{u_{i}}^{2}\bar{A}^{4}) f(M_{Q}^{i},M_{u}^{i},M_{d}^{i})\nonumber \\
&+&  f_{d_{i}}^{4}\bar{A}^{2}\bar{\mu}^{2} f(M_{Q}^{i},M_{d}^{i},M_{d}^{i}) 
+   f_{u_{i}}^{4}\bar{A}^{2}\bar{\mu}^{2} f(M_{Q}^{i},M_{u}^{i},M_{u}^{i}) ) ] 
\label{Lambda3bar}
\end{eqnarray}

\begin{eqnarray}
\bar{\Lambda}_{4} &=& \Lambda_{4}(1 - \sum_{i}{N_{c}f_{d_{i}}^{2}\bar{A}^{2}\over 12\pi(M_{Q}^{i} +
M_{d}^{i})^{3}} -\sum_{i} {N_{c}f_{u_{i}}^{2}\bar{\mu}^{2}\over 12\pi(M_{Q}^{i} +
M_{u}^{i})^{3}} \nonumber \\
&-& \sum_{i} {N_{c}f_{u_{i}}^{2}\bar{A}^{2}\over 12\pi(M_{Q}^{i} +
M_{u}^{i})^{3}} 
- \sum_{i}{N_{c}f_{d_{i}}^{2}\bar{\mu}^{2}\over 12\pi(M_{Q}^{i} +
M_{d}^{i})^{3}}) \nonumber \\
&+& \sum_{i} N_{c}[ \Lambda_{4}^{Q_{i}1}\Lambda_{4}^{Q_{i}2}{1\over 8\pi M_{Q}^{i}}
+ (\bar{f_{u_{i}}^{L}}^{2} \Lambda_{4}^{Q_{i}1} + 
\bar{f_{d_{i}}^{L}}^{2} \Lambda_{4}^{Q_{i}2}
+ 2\bar{f_{u_{i}}^{L}}^{2}\bar{f_{d_{i}}^{L}}^{2}){1\over 8\pi M_{Q}^{i}}  \nonumber \\
&+& (f_{u_{i}}^{2}\bar{\mu}^{2}(\bar{f_{u_{i}}^{L}}^{2}
+ \Lambda_{4}^{Q_{i}2}) - ( \bar{f_{d_{i}}^{L}}^{2} +\Lambda_{4}^{Q_{i}1})f_{u_{i}}^{2}\bar{A}^{2})
{1\over 8\pi M_{Q}^{i}}{1\over (M_{Q}^{i} + M_{u}^{i})^{2}} \nonumber \\
&+&  (f_{d_{i}}^{2}\bar{\mu}^{2}(\bar{f_{d_{i}}^{L}}^{2}
+ \Lambda_{4}^{Q_{i}1}) - (\bar{f_{u_{i}}^{L}}^{2}+ \Lambda_{4}^{Q_{i}2}) f_{d_{i}}^{2}\bar{A}^{2})
{1\over 8\pi M_{Q}^{i}}{1\over (M_{Q}^{i} + M_{d}^{i})^{2}}\nonumber \\
&-& 2( -( f_{d_{i}}^{2} f_{u_{i}}^{2}\bar{\mu}^{4}
-  2  f_{d_{i}}^{2} f_{u_{i}}^{2}\bar{A}^{2}\bar{\mu}^{2}
+  f_{d_{i}}^{2} f_{u_{i}}^{2}\bar{A}^{4})f(M_{Q}^{i},M_{u}^{i},M_{d}^{i})\nonumber \\ 
&+& 2 f_{u_{i}}^{4}\bar{A}^{2}\bar{\mu}^{2} f(M_{Q}^{i},M_{u}^{i},M_{u}^{i})
+  2 f_{d_{i}}^{4}\bar{A}^{2}\bar{\mu}^{2} 
 f(M_{Q}^{i},M_{u}^{i},M_{d}^{i}))]
\label{Lambda4bar}
\end{eqnarray}
where

\begin{equation}
f(m_{1},m_{2},m_{3}) = {1\over 8\pi} {2m_{1} + m_{2} + m_{3} \over
m_{1}(m_{1} + m_{3})^{2} (m_{1} + m_{2})^{2} (m_{2} + m_{3})}.
\end{equation}

\subsection{Two Light Higgses}

As mentioned in section 2.5, for the case in which both eigenvalues of the mass matrix of the Higgs doublets are such that
we cannot integrate out one of the scalar Higgs fields, the third stage corresponds
to the integration of only  the $A_{o}$ field. Since there is no trilinear $A_{o}\phi\phi$
interaction term there is no
wavefunction renormalization at this stage. Consequently, the two Higgs doublet potential is 

\begin{eqnarray}
V(A_{o},\phi_{1}, \phi_{2})& = & \bar{m}_{1}^{2} \phi_{1}^{\dagger}\phi_{1} + \bar{m}_{2}^{2}
 \phi_{2}^{\dagger}\phi_{2} 
+ \bar{m}_{3}^{2} (\phi_{1}^{\dagger}\phi_{2} + \phi_{2}^{\dagger}\phi_{1})
+ \bar{\lambda}_{1} (\phi_{1}^{\dagger}\phi_{1})^{2} \nonumber \\
& + &
\bar{\lambda}_{2} (\phi_{2}^{\dagger}\phi_{2})^{2} + \bar{\lambda}_{3}
 (\phi_{1}^{\dagger}\phi_{1})(\phi_{2}^{\dagger}\phi_{2}) +
\bar{\lambda}_{4} (\phi_{1}^{\dagger}\phi_{2})(\phi_{2}^{\dagger}\phi_{1}) ,
\label{third3dpot}
\end{eqnarray}
where

\begin{equation}
\bar{m}_{1}^{2} = \bar{M_{1}^{2}} - 3 \bar{H}_{1} {\bar{M}_{D}\over 4\pi} 
\label{barm12}
\end{equation}

\begin{equation}
\bar{m}_{2}^{2} = \bar{M_{2}^{2}} - 3 \bar{H}_{2} {\bar{M}_{D}\over 4\pi} 
\label{barm22}
\end{equation}

\begin{equation}
\bar{\lambda}_{1} = \bar{\Lambda}_{1} - 3 {\bar{H}_{1}^{2}\over 8\pi \bar{M}_{D}} 
\label{barlambda1}
\end{equation}

\begin{equation}
\bar{\lambda}_{2} = \bar{\Lambda}_{2} - 3 {\bar{H}_{2}^{2}\over 8\pi \bar{M}_{D}} 
\label{barlambda2}
\end{equation}

\begin{equation}
\bar{\lambda}_{3} = \bar{\Lambda}_{3} - 6 {\bar{H}_{1} \bar{H}_{2}\over 8\pi \bar{M}_{D}} 
\label{barlambda3}
\end{equation}
and $ \bar{m}_{3}^{2} =  \bar{M_{3}^{2}}$ and $\bar{\lambda}_{4} = \bar{\Lambda}_{4}$.

\section{2HDM and NMSSM}

\hspace*{2em} Our discussion of the 2HDM and NMSSN will be brief as
we have already introduced all of the relevant points in presenting the
effective theory for the MSSM. We will limit ourselves as much as possible to giving
 our results after each stage.

\subsection{Two Higgs Doublet Model}

\hspace*{2em} In the case of a general
two Higgs doublet model the scalar potential can contain additional quartic 
terms of the form,

\begin{equation}
\Delta V =\lambda_{5} (\phi_{1}^{\dagger}\phi_{2})(\phi_{1}^{\dagger}\phi_{2}) +
\lambda_{6} (\phi_{1}^{\dagger}\phi_{1})(\phi_{2}^{\dagger}\phi_{1})  +
\lambda_{7} (\phi_{2}^{\dagger}\phi_{2})(\phi_{2}^{\dagger}\phi_{1}) + h.c.
\label{2hdm}
\end{equation}
In this case the values of the $\lambda_{i}$ are not expressed in terms of
the weak coupling constant. We take all parameters to be real.

The reduction procedure differs from that of the MSSM  because the 
model does not contain superpartners. This implies that we will
have only two stages of reduction. The first one would once again correspond
to the integration out of the heavy non-static modes. Consequently, 
the $SU(3)$ gauge particles decouple when the fermions are eliminated. The resulting theory
for the static modes is be described by a scalar potential
with scalar masses

\begin{equation}
M_{D}^{2} = {g^2 T^2\over 6}(4 + N_{s} + N_{F}/2)
\label{MD2hdm}
\end{equation}

\begin{equation}
\Delta M_{1}^{2} = - 6 m_{3}^{2} \lambda_{6} {L_{b}\over 16\pi^2}
\end{equation}

\begin{equation}
\Delta M_{2}^{2} = - 6 m_{3}^{2} \lambda_{7} {L_{b}\over 16\pi^2}
\end{equation}

\begin{equation}
\Delta M_{3}^{2} = {T^{2}\over 4} (\lambda_{6} +  \lambda_{7})
- ( 12 \lambda_{5} m_{3}^{2} + 3 \lambda_{6} m_{1}^{2} + 3 \lambda_{7} m_{2}^{2} ) {L_{b}\over 16\pi^2}.
\end{equation}
We note that, unlike in the MSSM, the $M_{3}^{2}$ term receives a contribution
proportional to $T^{2}$, which is in fact the dominant correction. The quartic Higgs couplings are 
modified by the terms:

\begin{equation}
\Delta \Lambda_{1} = - T(2 \lambda_{5}^{2} + 6 \lambda_{6}^{2}) {L_{b}\over 16\pi^2}
\end{equation}

\begin{equation}
\Delta \Lambda_{2} = - T(2 \lambda_{5}^{2} + 6 \lambda_{7}^{2}) {L_{b}\over 16\pi^2}
\end{equation}

\begin{equation}
\Delta \Lambda_{3} = - T(4 \lambda_{5}^{2} + 2 \lambda_{6}^{2} + 8
\lambda_{6} \lambda_{7} +  2 \lambda_{7}^{2} ) {L_{b}\over 16\pi^2}
\end{equation}

 \begin{equation}
\Delta \Lambda_{4} = - T(32 \lambda_{5}^{2} + 5 \lambda_{6}^{2} + 2
\lambda_{6} \lambda_{7} +  5 \lambda_{7}^{2} ) {L_{b}\over 16\pi^2}
\end{equation}

\begin{eqnarray}
 \Lambda_{5}& = & \lambda_{5}T(1 + {9\over 2} g^2 {L_{b}\over 16\pi^2} - 
N_{c}\sum_{i}(f_{u_{i}}^{2} +  
f_{d_{i}}^2 {L_{f}\over 16\pi^2})
- T(4 \lambda_{1}\lambda_{5} + 4\lambda_{2}\lambda_{5}  \nonumber \\
&+&
8\lambda_{3}\lambda_{5} + 12\lambda_{4}\lambda_{5}
 + 5 \lambda_{6}^{2} + 2
\lambda_{6} \lambda_{7} +  5 \lambda_{7}^{2} ) {L_{b}\over 16\pi^2}
\end{eqnarray}

\begin{eqnarray}
 \Lambda_{6} &=&  \lambda_{6}T(1 + {9\over 2} g^2 {L_{b}\over 16\pi^2} - 
N_{c}\sum_{i}({f_{u_{i}}^{2}\over 2} +  
{3f_{d_{i}}^2\over 2} {L_{f}\over 16\pi^2}) 
- T(12 \lambda_{1}\lambda_{6} + 3\lambda_{3}\lambda_{6} \nonumber \\
&+&
4\lambda_{4}\lambda_{6} + 10\lambda_{5}\lambda_{6} + 3\lambda_{3}\lambda_{7} + 2\lambda_{4}\lambda_{7}
 + 2\lambda_{5} \lambda_{7}) {L_{b}\over 16\pi^2}
\end{eqnarray}

\begin{eqnarray}
 \Lambda_{7} &=&  \lambda_{7}T(1 + {9\over 2} g^2 {L_{b}\over 16\pi^2} - 
N_{c}\sum_{i}({f_{d_{i}}^{2}\over 2} +  
{3f_{u_{i}}^2\over 2} {L_{f}\over 16\pi^2}) 
- T(12 \lambda_{2}\lambda_{7} + 3\lambda_{3}\lambda_{7} \nonumber \\
&+&
4\lambda_{4}\lambda_{7} + 10\lambda_{5}\lambda_{7} + 3 \lambda_{3}\lambda_{6} + 2\lambda_{4}\lambda_{6}
  + 2
\lambda_{5} \lambda_{6}) {L_{b}\over 16\pi^2}.
\end{eqnarray}
We have written above  only the additional contributions but we remind
the reader that superpartner contributions to the formulae in
Appendix B must be dropped. This is true for the $G$ and $H$ couplings as well, which do not
receive new additional contributions from the extra interaction terms.

For the second stage there are two possibilities. First, as in 
the generic case of the MSSM, after the first stage one Higgs is much
heavier than the other and it can be integrated out with the $A_{o}$ field
after the mass matrix has been diagonalized. This is completely analogous to
the procedure in section  2.4, with the parameters changed 
as indicated above (ignoring all bars in the parameters
of section 2.4). The expressions for the $\alpha_{i}$ in equations (\ref{alpha1}),
(\ref{alpha3}) and (\ref{alpha4}) have additional contributions
from the $\Lambda_{5}$,  $\Lambda_{6}$ and $ \Lambda_{7}$ terms:

\begin{equation}
\Delta \alpha_{1} =  2\Lambda_{5} \cos^{2}\theta\sin^{2}\theta
+2\Lambda_{6}\cos^{4}\theta\sin\theta + \Lambda_{7}\sin^{3}\theta \cos\theta
\label{delalpha1}
\end{equation}

\begin{equation}
\Delta \alpha_{3} = -4\Lambda_{5}\cos^{2}\theta\sin^{2}\theta - 
(2\Lambda_{6}- 2\Lambda_{7})(\cos^{3}\theta\sin\theta - \sin^{3}\theta\cos\theta)
\label{delalpha3}
\end{equation}

\begin{equation}
\Delta \alpha_{4} = -4\Lambda_{5}\cos^{2}\theta\sin^{2}\theta - 
(2\Lambda_{6}- 2\Lambda_{7})(\cos^{3}\theta\sin\theta - \sin^{3}\theta\cos\theta)
\label{delalpha4}
\end{equation}
The second possibility is that both
Higgs fields are light in which case only the $A_{o}$ field is integrated out at this
second stage.
This is identical to the situation described in Appendix B.3. The quantities
 $\Lambda_{5}$,  $\Lambda_{6}$ and $ \Lambda_{7}$ are not modified by the $A_{o}$ field.

\subsection{Next to Minimal Supersymmetric Standard Model}

\hspace*{2em} If we now turn to the supersymmetric case with an additional 
 singlet superfield $\hat{N}$, we  will have additional terms 
 in the superpotential of the form \cite{haber, kane}

\begin{eqnarray}
\Delta W &=& \lambda(\hat{\phi_{1}^{o}}\hat{\phi_{2}^{o}} + 
\hat{\phi_{1}^{+}}\hat{\phi_{2}^{-}}) \hat{N} - {k\over 3} \hat{N}^{3} - r \hat{N}.
\label{deltaW}
\end{eqnarray}
Consequently, the extra terms in the scalar potential, including additional soft SUSY breaking
terms, are

\begin{eqnarray}
\Delta V &=& m_{N} N^{\ast}N + m_{4}\phi_{1}^{\dagger}\phi_{2} N + {1\over 3} m_{5}N^{3} + m_{7}^{2} N^{2}
+ \lambda_{5} (\phi_{1}^{\dagger}\phi_{1})(N^{\ast} N) \nonumber \\
&+&\lambda_{6} (\phi_{2}^{\dagger}\phi_{2})(N^{\ast} N)  +
\lambda_{7} (\phi_{1}^{\dagger}\phi_{2})N^{\ast^{2}}
+ \lambda_{8} (N^{\ast} N)^{2} 
+ \lambda_{9} (\phi_{1}^{\dagger}\phi_{2})(\phi_{2}^{\dagger}\phi_{1})\nonumber \\
& +& \lambda \mu N (\phi_{1}^{\dagger}\phi_{1} +
\phi_{2}^{\dagger}\phi_{2}) 
+ \lambda f_{d_{i}}\phi_{2}^{\dagger} Q_{i} D_{i}^{\ast} N^{\ast}
- \epsilon_{\alpha\beta} \lambda f_{u_{i}}\phi_{1}^{\alpha} Q_{i}^{\beta^{\ast}} U_{i}^{\ast} N^{\ast} + h.c.
\label{nmssm}
\end{eqnarray}
The quartic couplings are expressed in terms of the parameters in the superpotential
at the SUSY scale by

\begin{eqnarray}
\lambda_{5} &=& \lambda^{2}, \hspace{1cm} \lambda_{6} = \lambda^{2},  \hspace{1cm} \lambda_{9} = \lambda^{2} \\
\lambda_{7} &=& -\lambda k, \hspace{2cm} \lambda_{8} =  k^{2}.
\label{lambdanmssm}
\end{eqnarray}


We will have three reduction stages just as in the MSSM. For the first stage,
we see that the $G$ and $H$ couplings and the weak and strong Debye masses are not modified because the
particles  we have introduced
are gauge singlets. The 3D squarks masses are also not modified by the introduction of the
singlets. We mention that there are additional contributions to the 
wavefunction renormalization of the $\phi_{1}$ and $\phi_{2}$ fields from the
fermionic loops involving the singlet higgsino.
 For the scalar Higgs doublet masses and quartic couplings we have  additional contributions:

\begin{eqnarray}
\Delta M_{1}^{2} &=& (\lambda_{5} + \lambda_{9} + \lambda^{2}) {T^{2}\over 12}- m_{1}^{2}\lambda^{2} {L_{f}\over 16\pi^2}
 - ( m_{2}^{2} \lambda_{9} + m_{N}^{2} \lambda_{5}\nonumber\\
&+& m_{4}^{2} + 2\lambda^{2} \mu^{2}) {L_{b}\over 16\pi^2}
+ 2 \lambda^{2}\mu^{2} {L_{f}\over 16\pi^2}
\end{eqnarray}

\begin{eqnarray}
\Delta M_{2}^{2} &=&  (\lambda_{6} + \lambda_{9} + \lambda^{2}) {T^{2}\over 12} - m_{2}^{2}\lambda^{2} {L_{f}\over 16\pi^2}
- (m_{1}^{2}\lambda_{9} +  m_{N}^{2} \lambda_{6} \nonumber \\
&+& m_{4}^{2} + 2\lambda^{2} \mu^{2})  {L_{b}\over 16\pi^2}
+ 2 \lambda^{2}\mu^{2} {L_{f}\over 16\pi^2}
\end{eqnarray}

\begin{equation}
\Delta M_{3}^{2} = - m_{3}^{2}\lambda^{2} {L_{f}\over 16\pi^2} -(2  m_{3}^{2}\lambda_{9}
+  2 \lambda \mu m_{4} - 2 \lambda_{7} m_{7}^{2}) {L_{b}\over 16\pi^2}.
\end{equation}

\begin{equation}
\Delta \Lambda_{1} = - T\left(2\lambda_{1}\lambda^{2} {L_{f}\over 16\pi^2}+ 
({\lambda_{5}^{2}\over 2}  + {\lambda_{9}\over 2} + (\lambda_{3} + \lambda_{4})\lambda_{9}){L_{b}\over 16\pi^2}
 - \lambda^{4}{L_{f}\over 16\pi^2}\right)
\end{equation}

\begin{equation}
\Delta \Lambda_{2} = - T\left(2\lambda_{2}\lambda^{2} {L_{f}\over 16\pi^2}
+  ({\lambda_{6}^{2}\over 2} + {\lambda_{9}\over 2} + (\lambda_{3} + \lambda_{4})\lambda_{9}) {L_{b}\over 16\pi^2} - \lambda^{4}{L_{f}\over 16\pi^2}\right)
\end{equation}

\begin{eqnarray}
\Delta \Lambda_{3} &=& - T(2\lambda_{3}\lambda^{2} {L_{f}\over 16\pi^2}+ (\lambda_{5}\lambda_{6}
+ 2\lambda_{1}\lambda_{9} +   2\lambda_{2}\lambda_{9} +  2\lambda_{4}\lambda_{9} + \lambda_{9}^{2}) {L_{b}\over 16\pi^2})
\nonumber\\
&+& 2\lambda^{4}{L_{f}\over 16\pi^2}
\end{eqnarray}

\begin{eqnarray}
\Lambda_{4}& =  &(\lambda_{4} + \lambda_{9})T(1 + {9\over 2} g^2 {L_{b}\over 16\pi^2} - N_{c}\sum_{i}
(f_{u_{i}}^2 +
f_{d_{i}}^2) {L_{f}\over 16\pi^2}
- 3 g^2 {L_{f}\over 16\pi^2}  \nonumber \\
&-& 2 \lambda^{2}{L_{f}\over 16\pi^2})
+ T[-(2\lambda_{1}(\lambda_{4} + \lambda_{9})
+ 2\lambda_{2}(\lambda_{4} + \lambda_{9})
+ 2(\lambda_{4}+ \lambda_{9})^{2} \nonumber \\
&+& 4\lambda_{3}(\lambda_{4}+ \lambda_{9}) + 2 \lambda_{7}^{2})
{L_{b}\over 16\pi^2}
+ N_{c}\sum_{i} [\lambda_{4}(f_{d_{i}}^2 + f_{u_{i}}^2){L_{b}\over 16\pi^2}
 + 2 f_{u_{i}}^{2}f_{d_{i}}^{2}] {L_{b}\over 16\pi^2} \nonumber \\
&+& N_{sq}\lambda_{4}^{2}{L_{b}\over 16\pi^2} - 2N_{c}\sum_{i}
f_{u_{i}}^{2}f_{d_{i}}^{2}{L_{f}\over 16\pi^2}
- 2 g^4 {L_{f}\over 16\pi^2} ].
\label{lambda9}
\end{eqnarray}
The interaction terms between the scalar Higgs doublets and the squarks and
sleptons are modified by 

\begin{equation}
\Delta \Lambda_{3}^{Q_{i}1} = T(-\lambda_{3}\lambda^{2} {L_{f}\over 16\pi^2} - (\lambda^{2} f_{u_{i}}^{2}
+ \lambda_{9}( \lambda_{3} +  \lambda_{4} +  f_{u_{i}}^{2}) {L_{b}\over 16\pi^2}
+ 2 \lambda^{2} f_{u_{i}}^{2} {L_{f}\over 16\pi^2})
\end{equation}

\begin{equation}
\Delta \Lambda_{3}^{Q_{i}2} = T(-\lambda_{3}\lambda^{2} {L_{f}\over 16\pi^2} - (\lambda^{2} f_{d_{i}}^{2}
 + \lambda_{9}( \lambda_{3} +  \lambda_{4} +  f_{d_{i}}^{2}){L_{b}\over 16\pi^2}
+ 2 \lambda^{2} f_{d_{i}}^{2} {L_{f}\over 16\pi^2})
\end{equation}

\begin{equation}
\Delta (\bar{f_{d_{i}}^{L}}^{2} + \Lambda_{4}^{Q_{i}1}) = T( -(f_{d_{i}}^{2} + \lambda_{4})
\lambda^{2} {L_{f}\over 16\pi^2}
+(\lambda^{2}
 f_{u_{i}}^{2} + \lambda_{9}(f_{u_{i}}^{2} + \lambda_{4}) {L_{b}\over 16\pi^2}
- 2 \lambda^{2} f_{u_{i}}^{2} {L_{f}\over 16\pi^2})
\end{equation}

\begin{equation}
\Delta (\bar{f_{u_{i}}^{L}}^{2} + \Lambda_{4}^{Q_{i}2}) = T(-(f_{u_{i}}^{2} + \lambda_{4})
\lambda^{2} {L_{f}\over 16\pi^2}
+ (\lambda^{2}
 f_{d_{i}}^{2}  + \lambda_{9}(f_{d_{i}}^{2} + \lambda_{4}) {L_{b}\over 16\pi^2}
- 2 \lambda^{2} f_{d_{i}}^{2} {L_{f}\over 16\pi^2})
\end{equation}

\begin{equation}
\Delta \bar{f_{u_{i}}^{R }}^{2} = T( -f_{u_{i}}^{2} 
\lambda^{2} {L_{f}\over 16\pi^2}- 2\lambda^{2}
 f_{u_{i}}^{2} {L_{b}\over 16\pi^2}
+ 2 \lambda^{2} f_{u_{i}}^{2} {L_{f}\over 16\pi^2})
\end{equation}

\begin{equation}
\Delta \bar{f_{d_{i}}^{R }}^{2} = T(-f_{d_{i}}^{2} 
\lambda^{2} {L_{f}\over 16\pi^2}- 2\lambda^{2}
 f_{d_{i}}^{2} {L_{b}\over 16\pi^2}
+ 2 \lambda^{2} f_{d_{i}}^{2} {L_{f}\over 16\pi^2}).
\end{equation}

\begin{equation}
\Delta \bar{A}f_{u_{i}} = - T^{1/2}[A f_{u_{i}}{\lambda^{2}\over 2} {L_{f}\over 16\pi^2} + \lambda f_{u_{i}}
m_{4}  {L_{b}\over 16\pi^2}]
\end{equation}

\begin{equation}
\Delta \bar{A}f_{d_{i}} = - T^{1/2}[f_{d_{i}}{\lambda^{2}\over 2} {L_{f}\over 16\pi^2} + \lambda f_{d_{i}}
m_{4}  {L_{b}\over 16\pi^2}]
\end{equation}

\begin{equation}
\Delta \bar{\mu}f_{u_{i}} = - T^{1/2}[\mu f_{u_{i}}{\lambda^{2}\over 2} {L_{f}\over 16\pi^2} + \lambda^{2} f_{u_{i}}
\mu  {L_{b}\over 16\pi^2}]
\end{equation}

\begin{equation}
\Delta \bar{\mu}f_{d_{i}} = - T^{1/2}[\mu f_{d_{i}}{\lambda^{2}\over 2} {L_{f}\over 16\pi^2} + \lambda^{2} f_{d_{i}}
\mu  {L_{b}\over 16\pi^2}]
\end{equation}
The 3D expressions after the first stage for the mass and interaction
terms of the singlet Higgs are:

\begin{eqnarray}
\Lambda_{5} &=& \lambda_{5}T(1  + {9\over 4} g^2 {L_{b}\over 16\pi^2} - N_{c}\sum_{i} f_{d_{i}}^2
 {L_{f}\over 16\pi^2}
- {3\over 2}g^2 {L_{f}\over 16\pi^2} - (2k^{2} + 3\lambda^{2}){L_{f}\over 16\pi^2}) \nonumber \\
&-& T[(6\lambda_{1}\lambda_{5} + 2 \lambda_{5}^{2} + 2\lambda_{3}\lambda_{6} + 
\lambda_{4}\lambda_{6} + 4 \lambda_{7}^{2} + 4 \lambda_{5}\lambda_{8}
\nonumber \\
& +&  N_{c}\sum_{i} f_{u_{i}}^{2}
\lambda^{2}){L_{b}\over 16\pi^2} - (3g^{2}\lambda^{2} + 8 k^{2}\lambda^{2} + 2 \lambda^{4})
{L_{f}\over 16\pi^2}]
\end{eqnarray}

\begin{eqnarray}
\Lambda_{6} &=& \lambda_{6}T(1  + {9\over 4} g^2 {L_{b}\over 16\pi^2} - N_{c}\sum_{i} f_{u_{i}}^2
 {L_{f}\over 16\pi^2}
- {3\over 2}g^2 {L_{f}\over 16\pi^2}- (2k^{2} + 3\lambda^{2}){L_{f}\over 16\pi^2}) \nonumber \\
&-& T[(6\lambda_{2}\lambda_{6} + 2 \lambda_{6}^{2} + 2\lambda_{3}\lambda_{5} + 
\lambda_{4}\lambda_{5} + 4 \lambda_{7}^{2} + 4 \lambda_{6}\lambda_{8}
\nonumber \\
& +&  N_{c}\sum_{i} f_{d_{i}}^{2}
\lambda^{2}){L_{b}\over 16\pi^2} - (3g^{2}\lambda^{2} + 8 k^{2}\lambda^{2} + 2 \lambda^{4})
{L_{f}\over 16\pi^2}]
\end{eqnarray}

\begin{eqnarray}
\Lambda_{7} &=& \lambda_{7}T(1  + {9\over 4} g^2 {L_{b}\over 16\pi^2} - N_{c}\sum_{i}({f_{u_{i}}^2
\over 2} + {f_{d_{i}}^2\over 2}) {L_{f}\over 16\pi^2}
- {3\over 2}g^2 {L_{f}\over 16\pi^2}- (2k^{2} \nonumber \\
&+& 3\lambda^{2}){L_{f}\over 16\pi^2}) 
- T[(2\lambda_{6}\lambda_{7} + 2 \lambda_{7}\lambda_{8} + 2\lambda_{4}\lambda_{7} + 
2\lambda_{5}\lambda_{7} +  \lambda_{3}\lambda_{7}){L_{b}\over 16\pi^2}
\nonumber \\
&-&  2 k\lambda^{3}{L_{f}\over 16\pi^2}]
\end{eqnarray}

\begin{eqnarray}
\Lambda_{8} &=& \lambda_{8}T(1- 4(k^{2} + \lambda^{2}){L_{f}\over 16\pi^2})
- T[(\lambda_{5}^{2} + \lambda_{6}^{2} + 2\lambda_{7}^{2} +
10\lambda_{8}^{2})
{L_{b}\over 16\pi^2} \nonumber \\
&-& (8k^{4} + 2\lambda^{4}){L_{f}\over 16\pi^2}]
\end{eqnarray}

\begin{eqnarray}
M_{5} &=& m_{5}T^{1/2}(1- 3(k^{2} + \lambda^{2}){L_{f}\over 16\pi^2} - 3T^{1\over 2}[(2\lambda_{7} m_{4}
\nonumber \\
& +& 2\lambda_{8} m_{5} + 2\lambda \lambda_{6}\mu
+2 \lambda \lambda_{5}\mu){L_{b}\over 16\pi^2} - 4 \lambda^{3}\mu{L_{f}\over 16\pi^2}]
\end{eqnarray}

\begin{eqnarray}
M_{4} &=& m_{4}T^{1/2}(1  + {9\over 4} g^2 {L_{b}\over 16\pi^2} - N_{c}\sum_{i}({f_{u_{i}}^2
\over 2} + {f_{d_{i}}^2\over 2}) {L_{f}\over 16\pi^2}
- {3\over 2}g^2 {L_{f}\over 16\pi^2}-(k^{2}  \nonumber \\
&+& 2\lambda^{2}){L_{f}\over 16\pi^2})
-T^{1/2}[(4\lambda \lambda_{7}\mu + \lambda_{3}m_{4} + 2(\lambda_{4}+ \lambda_{9})m_{4} + \lambda_{5}m_{4}
+ \lambda_{6}m_{4} \nonumber \\
&+& 2\lambda_{7}m_{5}
+ A\lambda N_{c}\sum_{i}(f_{d_{i}}^{2} + f_{u_{i}}^{2})){L_{b}\over 16\pi^2}\nonumber \\
&-& (3g^{2}
\lambda m_{1/2} + 4 k \lambda^{2}\mu){L_{f}\over 16\pi^2}]
\end{eqnarray}

\begin{eqnarray}
M_{N} &=& m_{N}^{2}(1 - 2(k^2 + \lambda^{2}) {L_{f}\over 16\pi^2}) + (k^{2} + \lambda^{2}
+ \lambda_{5} + \lambda_{6} + 2\lambda_{8}){T^{2}\over 6} \nonumber \\
&-&  {L_{b}\over 16\pi^2}(8\lambda^{2}\mu^{2} + 2\lambda_{5}^{2} m_{1}^{2} +  2\lambda_{6}^{2} m_{2}^{2}
+  4\lambda_{7}^{2} m_{3}^{2} + 4 \lambda_{8}^{2} m_{N}^{2} + 2 m_{4}^{2} + 2 m_{5}^{2}
\nonumber \\
&-& 4 m_{7}^{2}\lambda_{8})
+ 12 \lambda^{2}\mu^{2} {L_{f}\over 16\pi^2}
\end{eqnarray}

\begin{eqnarray}
\Lambda f_{u_{i}} &=& \Lambda f_{u_{i}} T(1 +   {9\over 4} g^2 {L_{b}\over 16\pi^2}
 - ({N_{c}\over 2}\sum_{i} f_{d_{i}}^2 + {1\over 2} f_{d_{i}}^2 +
{3\over 2}f_{u_{i}}^{2}) {L_{f}\over 16\pi^2}
- {3\over 2}g^2 {L_{f}\over 16\pi^2} \nonumber \\
&+& 4g_{s}^{2} {L_{b}\over 16\pi^2} - {8\over 3}g_{s}^{2} {L_{f}\over 16\pi^2}
- ({3\over 2} \lambda^{2} + k^{2}) {L_{f}\over 16\pi^2}) - T[ {L_{b}\over 16\pi^2}
(-2 \lambda f_{u_{i}}  f_{d_{i}}^{2} \nonumber \\
&-& {4\over 3} \lambda f_{u_{i}} g_{s}^{2}  
+ \lambda f_{u_{i}} \lambda_{3} + \lambda f_{u_{i}} \lambda_{5} - \lambda f_{u_{i}} \lambda_{4}
+ N_{c}\lambda f_{u_{i}}\sum_{i} f_{u_{i}}^{2}) \nonumber \\
&-& {L_{f}\over 16\pi^2} (- 2\lambda f_{u_{i}} 
 f_{d_{i}}^{2} + 3 \lambda f_{u_{i}} g^{2})]
\end{eqnarray}

\begin{eqnarray}
\Lambda f_{d_{i}} &=& \Lambda f_{d_{i}} T(1 +   {9\over 4} g^2 {L_{b}\over 16\pi^2}
 - ({N_{c}\over 2}\sum_{i} f_{u_{i}}^2 + {1\over 2} f_{u_{i}}^2 +
{3\over 2}f_{d_{i}}^{2}) {L_{f}\over 16\pi^2}
- {3\over 2}g^2 {L_{f}\over 16\pi^2} \nonumber \\
&+& 4g_{s}^{2} {L_{b}\over 16\pi^2} - {8\over 3}g_{s}^{2} {L_{f}\over 16\pi^2}
- ({3\over 2} \lambda^{2} + k^{2}) {L_{f}\over 16\pi^2}) - T[ {L_{b}\over 16\pi^2}
(-2 \lambda f_{d_{i}}  f_{u_{i}}^{2}  \nonumber \\
&-& {4\over 3} \lambda f_{d_{i}} g_{s}^{2} 
+ \lambda f_{d_{i}} \lambda_{3} + \lambda f_{d_{i}} \lambda_{6} - \lambda f_{d_{i}} \lambda_{4}
+ N_{c}\lambda f_{d_{i}}\sum_{i} f_{d_{i}}^{2})  \nonumber \\
&-& {L_{f}\over 16\pi^2} (- 2\lambda f_{d_{i}} 
 f_{u_{i}}^{2} + 3 \lambda f_{d_{i}} g^{2})].
\end{eqnarray}
We denote the 3D coupling of the trilinear $\phi_{1}^{\dagger}\phi_{1}N$ term  by $J_{1}$
and similarly for the $\phi_{2}^{\dagger}\phi_{2}N$ term  by $J_{2}$:

\begin{eqnarray}
J_{1} &=& \lambda\mu T^{1/2}(1  + {9\over 4} g^2 {L_{b}\over 16\pi^2} - N_{c}\sum_{i} f_{d_{i}}^2
 {L_{f}\over 16\pi^2}
- {3\over 2}g^2 {L_{f}\over 16\pi^2}-(k^{2} + 2\lambda^{2}){L_{f}\over 16\pi^2}) \nonumber \\
&-&T^{1/2}[(6 \lambda \lambda_{1}\mu + 2 \lambda \lambda_{3}\mu + \lambda \lambda_{4}\mu
+ 2\lambda \lambda_{5}\mu + N_{c}\sum_{i}\lambda f_{u_{i}}^{2}\mu) {L_{b}\over 16\pi^2}\nonumber \\
&-& (3g^{2} \lambda\mu + 2\lambda^{3}\mu){L_{f}\over 16\pi^2}]
\end{eqnarray}

\begin{eqnarray}
J_{2} &=& \lambda\mu T^{1/2}(1  + {9\over 4} g^2 {L_{b}\over 16\pi^2} - N_{c}\sum_{i} f_{u_{i}}^2
 {L_{f}\over 16\pi^2}
- {3\over 2}g^2 {L_{f}\over 16\pi^2}-(k^{2} + 2\lambda^{2}){L_{f}\over 16\pi^2}) \nonumber \\
&-&T^{1/2}[(6 \lambda \lambda_{2}\mu + 2 \lambda \lambda_{3}\mu + \lambda \lambda_{4}\mu
+ 2\lambda \lambda_{6}\mu + N_{c}\sum_{i}\lambda f_{d_{i}}^{2}\mu) {L_{b}\over 16\pi^2}\nonumber \\
&-& (3g^{2} \lambda\mu + 2\lambda^{3}\mu){L_{f}\over 16\pi^2}].
\end{eqnarray}
The second stage proceeds just like in the MSSM.
The interaction terms between doublet and singlet Higgs fields are modified by:

\begin{eqnarray}
\bar{\Lambda}_{5} &=& \Lambda_{5}(1 -  \sum_{i}{N_{c}f_{d_{i}}^{2}\bar{A}^{2}\over 12\pi(M_{Q}^{i} +
M_{d}^{i})^{3}} - \sum_{i}{N_{c}f_{u_{i}}^{2}\bar{\mu}^{2}\over 12\pi(M_{Q}^{i} +
M_{u}^{i})^{3}}) \nonumber \\
& -& \sum_{i}{\Lambda^{2} f_{u_{i}}^{2}\over
4\pi(M_{Q}^{i} + M_{u}^{i})}
\end{eqnarray}

\begin{eqnarray}
\bar{\Lambda}_{6} &=& \Lambda_{6}(1 -  \sum_{i}{N_{c}f_{u_{i}}^{2}\bar{A}^{2}\over 12\pi(M_{Q}^{i} +
M_{u}^{i})^{3}} - \sum_{i}{N_{c}f_{d_{i}}^{2}\bar{\mu}^{2}\over 12\pi(M_{Q}^{i} +
M_{d}^{i})^{3}}) \nonumber \\
& -& \sum_{i}{\Lambda^{2} f_{d_{i}}^{2}\over
4\pi(M_{Q}^{i} + M_{d}^{i})}
\end{eqnarray}

\begin{eqnarray}
\bar{M}_{4} &=& M_{4}(1 -  \sum_{i}{N_{c}f_{u_{i}}^{2}\bar{A}^{2}\over 24\pi(M_{Q}^{i} +
M_{u}^{i})^{3}} - \sum_{i}{N_{c}f_{d_{i}}^{2}\bar{\mu}^{2}\over 24\pi(M_{Q}^{i} +
M_{d}^{i})^{3}} \nonumber \\
&-& \sum_{i}{N_{c}f_{d_{i}}^{2}\bar{A}^{2}\over 24\pi(M_{Q}^{i} +
M_{d}^{i})^{3}} - \sum_{i} {N_{c}f_{u_{i}}^{2}\bar{\mu}^{2}\over 24\pi(M_{Q}^{i} +
M_{u}^{i})^{3}} ) \nonumber \\
& -& \sum_{i} N_{c}({\Lambda \bar{A}f_{d_{i}}^{2}\over
4\pi(M_{Q}^{i} + M_{d}^{i})} + {\Lambda \bar{A}f_{u_{i}}^{2}\over
4\pi(M_{Q}^{i} + M_{u}^{i})})
\end{eqnarray}

\begin{eqnarray}
\bar{J}_{1} &=& J_{1}(1 -  \sum_{i}{N_{c}f_{d_{i}}^{2}\bar{A}^{2}\over 12\pi(M_{Q}^{i} +
M_{d}^{i})^{3}} - \sum_{i}{N_{c}f_{u_{i}}^{2}\bar{\mu}^{2}\over 12\pi(M_{Q}^{i} +
M_{u}^{i})^{3}}) \nonumber \\
&-& N_{c}\sum_{i} {\Lambda\bar{\mu}f_{u_{i}}^{2}\over 4\pi(M_{Q}^{i} + M_{u}^{i})}
\end{eqnarray}

\begin{eqnarray}
\bar{J}_{2} &=& J_{2}(1 - \sum_{i}{N_{c}f_{u_{i}}^{2}\bar{A}^{2}\over 12\pi(M_{Q}^{i} +
M_{d}^{i})^{3}} +
- \sum_{i}{N_{c}f_{d_{i}}^{2}\bar{\mu}^{2}\over 12\pi(M_{Q}^{i} +
M_{u}^{i})^{3}}) \nonumber \\
&-& N_{c}\sum_{i} {\Lambda\bar{\mu}f_{d_{i}}^{2}\over 4\pi(M_{Q}^{i} + M_{d}^{i})}
\end{eqnarray}
There are no triangle and box diagram corrections to the above second stage quantities.

 The scalars in the resulting theory are
two Higgs doublets, a Higgs singlet and the $A_{o}$ triplet.
Depending on the values
of the parameters  which determine the 3D mass of the singlet, if it is
heavy ($\sim gT$) it can be integrated
out at the third stage after the diagonalization of the scalar doublets mass matrix.
We remind the reader that we have determined the critical temperature 
by finding the direction in which the curvature of the potential vanishes
at the origin. Consequently,  there are no mixing terms in the mass matrix between the doublets and the 
singlet Higgs fields. The additional contributions to $\bar{\lambda}_{3}$ from
the singlet,  including wavefunction corrections,  are:

\begin{eqnarray}
\Delta \bar{\lambda}_{3} &=& -\alpha_{1}({\alpha_{5}^{2} + \alpha_{6}^{2}
\over 6\pi(\nu(T_{c}) + M_{N})^{3}})- {\alpha_{7}^{2}\over 2} {1\over 8\pi M_{N}} 
-2 \Lambda_{7}^{2}\cos^{2}\theta\sin^{2}\theta {1\over 8\pi M_{N}} \nonumber \\
&+& \alpha_{7}(\alpha_{5}^{2} + \alpha_{6}^{2}){1\over 8\pi \nu(T_{c})}{1\over (\nu(T_{c}) + M_{N})^{2}}
\nonumber \\
&-& (\alpha_{5}^{2} + \alpha_{6}^{2})^{2}{1\over 8\pi \nu(T_{c}) M_{N}}{1\over (\nu(T_{c}) + M_{N})^{3}}
\end{eqnarray}
where

\begin{equation}
\alpha_{5} = \bar{M}_{4}\cos^{2}\theta - \bar{J}_{1}\sin\theta\cos\theta + \bar{J}_{2}\sin\theta\cos\theta
\end{equation}

\begin{equation}
\alpha_{6} = -\bar{M}_{4}\sin^{2}\theta - \bar{J}_{1}\sin\theta\cos\theta + \bar{J}_{2}\sin\theta\cos\theta
\end{equation}

\begin{equation}
\alpha_{7} = \bar{\Lambda}_{5}\cos^{2}\theta + \bar{\Lambda}_{6}\sin^{2}\theta.
\end{equation}


\newcommand{\Tint}{{\hbox{$\sum$}\!\!\!\!\!\!\int}}
\section{Appendix of Finite Temperature Formulae}

\hspace*{2em} In this appendix we include the basic integrals which appear in the calculation
over the non-static modes. These results can be derived from formulae presented in the literature.
 We refer the reader to references \cite{kajantie, moss, arnold}
 and references within
for more details regarding finite temperature formulae. 

\subsection{ $m = 0$.}

Let us consider first the massless case. We can define the following quantities

\begin{equation}
A_{s} = \Tint{1\over p^{2s}} = 2\mu^{2\epsilon} T {\Gamma(-{D\over2} + s)\over
(4\pi)^{D/2} \Gamma(s)} (2\pi T)^{-2s + D} \zeta(2s -D)
\label{aj}
\end{equation}
with $p^{2} = \vec{p}^{2} + \omega_{B} ^2$, $\omega_{B} = 2\pi n T$, $D= 3-2\epsilon$.
For bosonic sums, $n=0$ is excluded. Where,

\begin{equation}
\Tint = \mu^{2\epsilon}T\sum_{p_o}\int {d^{3-2\epsilon}p\over (2\pi)^{3-2\epsilon}}
\label{int}
\end{equation}
Similarly we have for fermionic excitations,

\begin{equation}
B_{j} = \Tint {1\over \tilde{p}^{2s}}
\label{bj}
\end{equation}
with $\tilde{p}^{2} = \vec{p}^{2} + \omega_{F}^2$, $\omega_{F} = 2\pi (n +1/2)T$.
Using the fact that 

\begin{equation}
A_{s} + B_{s} = 2^{2s -D} A_{s}
\label{ajbj}
\end{equation}
we can easily determine the fermionic contributions in terms of
the bosonic integrals.
Generalizing, we can write

\begin{equation}
B_{s}^{\alpha_{1}\cdots \alpha_{k}} = (2^{2s -D -k}-1)
 A_{s}^{\alpha_{1}\cdots \alpha_{k}}
\label{ajalpha}
\end{equation}
where the superscripts $\alpha_{1}\cdots \alpha_{k}$ indicate additional powers of
momenta in the integrals.
In particular we have:

\begin{equation}
A_{1} = {T^{2} \over 12} + {\cal O} (\epsilon)
\label{a1}
\end{equation}
and

\begin{equation}
A_{2} = {1\over (4\pi)^{2}}({1\over\epsilon} + L_{b})
\label{a2}
\end{equation}
where

\begin{eqnarray}
L_{b} & = & 2\log {\bar{\mu_{4}} e^{\gamma} \over 4\pi T} \nonumber \\
& = & -\log 4\pi T^{2} + \gamma + \log\mu_{4}^{2}.
\label{ls}
\end{eqnarray}
$\bar{\mu_{4}}$ is defined by the $\overline{MS}$ scheme.

\subsection{ $m\neq 0$}

If we now include the effect of masses our formulas will be modified in the
following way.

\begin{eqnarray}
A_{s}(m) &=& \Tint {1\over (p^{2} + m^{2})^{s}} = \mu^{2\epsilon}T {\Gamma(-{D\over2} +
s)\over
(4\pi)^{D/2} \Gamma (s)} (2\pi T)^{-2s + D} \nonumber \\
&\times & \left[\zeta(2s -D; {m\over 2\pi T}) 
- \left({m\over 2\pi T}\right)^{(-2s + D)} \right]
\label{asm}
\end{eqnarray}
where

\begin{equation}
\zeta(\sigma;\nu) = \sum_{n=-\infty}^{n=\infty} (n^{2} + \nu^{2})^{-\sigma}.
\label{zeta}
\end{equation}
It is easy to verify that

\begin{equation}
A_{s}(m) = -(s-1)^{-1} {\partial\over \partial m^{2}} A_{s-1}(m)
\label{aj1m}
\end{equation}
and  for high temperature, dropping the ${\cal O} (\epsilon)$ terms,
 we can use the expansion

\begin{equation}
A_{1}(m) = {T^{2} \over 12} - {m^{2}\over 16\pi^{2}} 
({1\over\epsilon} + L_{b}) + {\cal O}({m^4\over T^2})
\label{a1m}
\end{equation}
and

\begin{equation}
A_{2}(m) = {1\over (4\pi)^{2}}({1\over\epsilon} + L_{b})+ \cdots
\label{a2m}
\end{equation}
We can now extend our considerations to express the fermionic
integrals in terms of the bosonic ones, obtaining,

\begin{equation}
B_{s}^{\alpha_{1}\cdots \alpha_{k}} (m) = 2^{2s -D -k}
 A_{s}^{\alpha_{1}\cdots \alpha_{k}}(2m) - 
A_{s}^{\alpha_{1}\cdots \alpha_{k}}(m).
\label{ajalpham}
\end{equation}
Let us write the explicit results for:

\begin{equation}
A_s^{i} = \Tint{p_{i}\over (p^{2} + m^{2})^{s}} = 0
\label{kiaj}
\end{equation}

\begin{equation}
A_s^{0} =\Tint {p_{0}\over (p^{2} + m^{2})^{s}} = 0
\label{k0aj}
\end{equation}

\begin{eqnarray}
A_s^{ij} &= &\Tint{p_{i}p_{j}\over (p^{2} + m^{2})^{s}} = A_{1}(s)\delta_{ij}
\nonumber \\
A_{s}^{00} & = &\Tint {p_{0}p_{0}\over (p^{2} + m^{2})^{s}} = A_{1}(s) + A_{2}(s)
\label{kijaj}
\end{eqnarray}
where

\begin{eqnarray}
A_{i} (s) &= & -(s-1)^{-1} {\partial\over \partial m^{2}} A_{i} (s-1)
\nonumber \\
A_{1}(1) & =& {m^{4}\over 64\pi^{2}} ({1\over \epsilon} + L_{b})
+ {1\over 24} m^{2} T^{2} \nonumber \\
A_{2}(1) & = & -{2\pi^{2}\over 45} T^{4} + {1\over 32\pi^{2}} m^{4}
+ {1\over 12} m^{2}T^{2} + \cdots
\label{a1k}
\end{eqnarray}
We would also like to write the explicit results for the following
integrals,

\begin{eqnarray}
C(p, m_{1}, m_{2}) & = &\Tint{1\over k^{2} + \omega_{n}^{2} +m_1^2}{1\over (k-p)^{2}
 + \omega_{n}^{2} + m_2^2} \nonumber \\
&=& A_{2}
\label{c2}
\end{eqnarray}

\begin{eqnarray}
C^{i}(p,m_{1}, m_{2}) & = &\Tint{k_{i}\over k^{2} + \omega_{n}^{2} + m_1^2}{1\over
 (k-p)^{2} + \omega_{n}^{2}  + m_2^2} \nonumber \\
&=& p_{i}A_{2}/2
\label{c2i}
\end{eqnarray}

\begin{eqnarray}
C^{ij}(p,m_{1}, m_{2}) & = &\Tint{k_{i}k_{j}\over k^{2} + \omega_{n}^{2}
+ m_1^2}{1\over
 (k-p)^{2} + \omega_{n}^{2}  + m_2^2} \nonumber \\
&=& p_{i}p_{j} C_{21} + g_{ij}C_{22}
\label{c2ij}
\end{eqnarray}
with

\begin{equation}
C_{21} =  A_{2}/3\
\label{c21}
\end{equation}

\begin{equation}
C_{22} = {T^{2} \over 24} - (m_{1}^{2} + m_{2}^{2} + {p^{2}\over 3}) A_{2}/4
\label{c22}
\end{equation}

\begin{eqnarray}
C^{oo}(p,m_{1}, m_{2}) & = &\Tint{k_{o}^2\over k^{2} + \omega_{n}^{2}
+m_1^2}{1\over
 (k-p)^{2} + \omega_{n}^{2}  + m_2^2} \nonumber \\
& = & -{T^{2} \over 24} - {1\over 64\pi^2} ({1\over\epsilon} + L_{b} +2)(m_{1}^{2}
 + m_{2}^{2} + {p^{2}\over 3})
\label{coo}
\end{eqnarray}

\begin{eqnarray}
E^{ij}(p,m_{1}, m_{2}) & = &\Tint{k_{i}k_{j}\over k^{2} + \omega_{n}^{2}
+ m_1^2}{1\over
 (k-p)^{2} + \omega_{n}^{2}  + m_2^2} {1\over  (k-p')^{2} + \omega_{n}^{2}  + m_3^2}\nonumber \\
&=& \delta_{ij} {1\over 4}{1\over 16\pi^2} ({1\over\epsilon} + L_{b})
\label{e3ij}
\end{eqnarray}

\begin{eqnarray}
E^{oo}(p,m_{1}, m_{2}) & = &\Tint{k_{o}^2\over k^{2} + \omega_{n}^{2}
+m_1^2}{1\over
 (k-p)^{2} + \omega_{n}^{2}  + m_2^2} {1\over  (k-p')^{2} + \omega_{n}^{2}  + m_3^2}\nonumber \\
&=&  {1\over 16\pi^2}[{1\over 4} ({1\over \epsilon} + L_{b}) + {1\over 2}].
\label{e300}
\end{eqnarray}
We would like to relate the fermionic integrals of this type to the bosonic
ones.

\begin{equation}
D^{\alpha_{1}\cdots \alpha_{m}} (p) = \Tint{\tilde{k}_{\alpha_{1}}\cdots
\tilde{k}_{\alpha_{m}}\over \tilde{k}^{2} + \omega_{n}^{2}}{1\over
(\tilde{k}-\tilde{p})^{2} + \omega_{n}^{2}}
\label{dj}
\end{equation}
with, once again, $\tilde{p}^{2} = \vec{p}^{2} + \omega_{F}^2$, $\omega_{F} = 2\pi (n +1/2)T$.
 We obtain,

\begin{eqnarray}
D(p, m_{1}, m_{2}) & = &\Tint{1\over k^{2} + \omega_{n}^{2} +m_1^2}{1\over (k-p)^{2}
 + \omega_{n}^{2} + m_2^2} \nonumber \\
&=& B_{2}
\label{d2}
\end{eqnarray}

\begin{eqnarray}
D^{i}(p,m_{1}, m_{2}) & = &\Tint{k_{i}\over k^{2} + \omega_{n}^{2} + m_1^2}{1\over
 (k-p)^{2} + \omega_{n}^{2}  + m_2^2} \nonumber \\
&=& p_{i}B_{2}/2
\label{D2i}
\end{eqnarray}

\begin{eqnarray}
D^{ij}(p,m_{1}, m_{2}) & = &\Tint{k_{i}k_{j}\over k^{2} + \omega_{n}^{2}
+ m_1^2}{1\over
 (k-p)^{2} + \omega_{n}^{2}  + m_2^2} \nonumber \\
&=& p_{i}p_{j} D_{21} + g_{ij}D_{22}
\label{d2ij}
\end{eqnarray}
explicitly,

\begin{equation}
D_{21} =  B_{2}/3
\label{d21}
\end{equation}

\begin{equation}
D_{22} = -{T^{2} \over 48} - (m_{1}^{2} + m_{2}^{2} + {p^{2}\over 3}) B_{2}/4
\label{d22}
\end{equation}

\begin{eqnarray}
D^{oo}(p,m_{1}, m_{2}) & = &\Tint{k_{o}^2\over k^{2} + \omega_{n}^{2}
+ m_1^2}{1\over
 (k-p)^{2} + \omega_{n}^{2}  + m_2^2} \nonumber \\
&=& {T^{2} \over 48} - {1\over 64\pi^2} ({1\over\epsilon} + L_{f} +2)(m_{1}^{2}
 + m_{2}^{2} + {p^{2}\over 3})
\label{doo}
\end{eqnarray}
where $L_{f} = L_{b} + 4\ln 2$.





\end{document}